\documentclass[conference,compsoc]{IEEEtran}
\usepackage[]{mdframed}
\usepackage{etoolbox}
\usepackage{xurlshawn}
\usepackage[numbers,sort]{natbib}
\usepackage{balance}
\usepackage{url}
\usepackage{color}
\usepackage{caption}
\usepackage{diagbox}
\usepackage{subfigure}
\usepackage{algorithm,algpseudocode}
\usepackage{amsmath,amssymb}

\usepackage{multirow}
\usepackage{booktabs}
\usepackage{epsfig,endnotes}
\usepackage{enumitem}
\usepackage[font=footnotesize,labelfont=bf, labelsep=period]{caption}
\newcommand{\para}[1]{{\vspace{2pt} \noindent \textbf{#1}
    \hspace{3pt}}}

\newcommand{\shawn}[1]{{\color{black} #1}}
\newcommand{\shawnsp}[1]{{\color{black} #1}}
\newcommand{\final}[1]{{\color{black} #1}}

\newcommand{\htnew}[1]{{\color{black} #1}}

\newcommand{\lookatme}[1]{{\color{black} #1}}

\newcommand{\eg}{{\em e.g.,\ }}
\newcommand{\ie}{{\em i.e.,\ }}

\newcommand{\secspace}{\vspace{0.0in}}

\newcommand{\viccon}{\mathcal{C}}
\newcommand{\aidcon}{\mathcal{A}}

\newcommand{\dfunc}{D}

\newcommand{\framedtext}[1]{%
\par%
\noindent\fbox{%
    \parbox{\dimexpr\linewidth-2\fboxsep-2\fboxrule}{#1}%
}%
}

\newenvironment{packed_itemize}{
\begin{list}{\labelitemi}{\leftmargin=0.5em}
  \setlength{\itemsep}{3pt}
  \setlength{\parskip}{0pt}
  \setlength{\parsep}{0pt}
  \setlength{\headsep}{0pt}
  \setlength{\topskip}{0pt}
  \setlength{\topmargin}{0pt}
  \setlength{\topsep}{0pt}
  \setlength{\partopsep}{0pt}
}{\end{list}}

\newenvironment{packed_enumerate}{
\begin{enumerate}
 \setlength{\itemsep}{1pt}
 \setlength{\parskip}{0pt}
 \setlength{\parsep}{0pt}
 \setlength{\headsep}{0pt}
 \setlength{\topskip}{0pt}
 \setlength{\topmargin}{0pt}
 \setlength{\topsep}{0pt}
 \setlength{\partopsep}{0pt}
}{\end{enumerate}}

\begin{document}

\title{Nightshade: Prompt-Specific Poisoning Attacks on Text-to-Image Generative Models}
\author{Shawn Shan, Wenxin Ding, Josephine Passananti, Stanley Wu, Haitao Zheng, Ben Y. Zhao\\
  {\em Department of Computer Science, University of Chicago}\\
  {\em \{shawnshan, wenxind, josephinep, stanleywu, htzheng, ravenben\}@cs.uchicago.edu}}

\maketitle

\begin{abstract}
  Trained on billions of images,
  diffusion-based text-to-image models seem impervious to traditional data poisoning attacks, which
  typically require poison samples approaching 20\% of the training set.  
  In this paper, we show that state-of-the-art
  text-to-image generative models are in fact highly vulnerable to poisoning
  attacks. Our work is driven by two key insights. First, while diffusion
  models are trained on billions of samples, the number of
  training samples associated with a specific concept or prompt is generally on
  the order of thousands. This suggests that these models will be vulnerable
  to {\em prompt-specific poisoning attacks} that corrupt a model's ability
  to respond to specific targeted prompts. Second, poison samples can be
  carefully crafted to maximize poison potency to ensure success with very
  few samples.  
  
  We introduce {\em Nightshade}, a prompt-specific poisoning attack optimized for potency that
  can completely control the output of a prompt in Stable Diffusion's newest
  model (SDXL) with less than $100$ poisoned training samples.  Nightshade also
  generates stealthy poison images that look visually identical to their benign
  counterparts, and produces poison effects that ``bleed through'' to related
  concepts.  More importantly, a moderate number of Nightshade attacks on
  independent prompts can destabilize a model and disable its ability to
  generate images for any and all prompts. Finally, we propose the use of
  Nightshade and similar tools as a  defense for content owners against
  web scrapers that ignore opt-out/do-not-crawl directives, and discuss
  potential implications for both model trainers and content owners.

\end{abstract}

\section{Introduction}
\vspace{-0.05in}

Since 2022, diffusion based text-to-image models have taken the
Internet by storm, growing from research projects to numerous applications in
advertising, fashion~\cite{tryon,tryon2}, web
development~\cite{webdev1,webdev2,webdev3}, and AI
art~\cite{hollie-steal,sarah-andersen,lensa-steal,sam-steal}. Models like
Stable Diffusion SDXL, Midjourney v5, Dalle-3, Imagen, Adobe Firefly and others boast
tens of millions of registered users and have produced billions of images~\cite{adobemax}.

To date, public consensus considers these diffusion models impervious to 
data poisoning attacks. Poisoning attacks manipulate training data to
introduce unexpected behavior to the model at training time, and have been
studied extensively in the context of classification tasks using deep neural
networks (DNN).  Poisoning attacks cause predictable misclassifications, but
typically demand a substantial volume of poison data for success, e.g., ratio
of poison training samples to benign samples of 20\% or higher.  
Since today's diffusion models are trained on hundreds of millions (or billions) of
images, a common assumption is that poisoning attacks on these models would
require millions of poison samples, making them infeasible in
  practice.

In this work, we demonstrate a surprising result: state-of-the-art
text-to-image models are in fact highly vulnerable to data
poisoning attacks. Our work is based on two key insights. First, while
these  models are trained on millions and billions of images, the number of
  training samples associated with a specific concept or prompt is quite low, on
  the order of thousands. We call this property ``concept sparsity,'' and it
  suggests the viability of {\em prompt-specific poisoning
    attacks} that corrupt a model's ability to respond to specific targeted
  prompts. Second, we observe that natural benign images exhibit large
  variance in text labels, image composition, and image features, all of
  which produce destructive interference to minimize training influence. By
  crafting poison samples that minimize these sources of interference, we can
  produce highly effective poison attacks with very few samples. 
  Unlike previous work on backdoor
  attacks~\cite{chen2023trojdiff,chou2023backdoor,zhai2023text}, we show that
  successful prompt-specific poisoning attacks {\em do not} require access to
  the model internal pipeline, and only need a very small number of poison
  samples to override a specific target prompt. For example, a single Nightshade
  attack (``car'' to ``cow'') targeting Stable Diffusion SDXL has a high probability of success
  using only 50 optimized samples, and the poisoned model outputs an image
  of a cow for every mention of a car in its prompts.

\htnew{This paper describes our experiences and findings
  in designing and evaluating prompt-specific poisoning attacks
  against generative text-to-image models.  {\em First}, we 
  validate our hypothesis of ``concept sparsity'' in existing large-scale
  datasets used to train generative image models.}
We find that as hypothesized, concepts in popular training datasets like
LAION-Aesthetic exhibit very low training data density, both in terms of
concept sparsity (\# of training samples associated explicitly with a
specific concept) and semantic sparsity (\# of samples associated with a
concept and its semantically related terms).  {\em Second}, we confirm a
proof of concept poisoning attack (by mislabeling images) can successfully corrupt image
generation for specific concepts (\eg ``dog'') using 500-1000 poison
samples. Successful attacks on Stable Diffusion's newest model
(SDXL) are confirmed using both CLIP-based classification and an (IRB-approved)
user study. Unfortunately this attack still requires too many poison samples
and is easily detected/filtered.

{\em Third}, we propose a highly optimized prompt-specific poisoning attack
we call {\em Nightshade}. Nightshade uses multiple optimization techniques
\lookatme{(including targeted adversarial perturbations) to generate stealthy
  and highly effective} poison samples, with four observable benefits.

\begin{packed_enumerate}
\item Nightshade poison samples are benign images shifted in the feature
  space, and still look like their benign counterparts to the human eye. They
  avoid detection through human inspection and prompt generation.
\item Nightshade samples produce stronger poisoning effects, \lookatme{enabling} highly
  successful poisoning attacks with very few (\eg  \shawn{100}) samples. 
\item Nightshade's poisoning effects ``bleed through'' to related concepts, and
  thus cannot be circumvented by prompt replacement. For example,
  \lookatme{Nightshade samples} poisoning ``fantasy art'' also affect ``dragon''
  and ``Michael Whelan'' (a well-known fantasy and SciFi artist). Nightshade
  attacks are composable, e.g. a single prompt can trigger multiple poisoned
  prompts. 
\item  When many independent Nightshade attacks affect different prompts
    on a single model (\eg 250 attacks on SDXL), the model's understanding of
    basic features becomes corrupted and it is no longer able to generate 
    meaningful images.
\end{packed_enumerate}

\noindent We also observe \htnew{that Nightshade exhibits strong
  transferability across models and can resist a spectrum of defenses
  intended to deter current poisoning attacks.}

{\em Finally}, we propose the use of Nightshade as a powerful tool for content
owners to protect their intellectual property. Today, content owners can only
rely on opt-out lists and do-not-scrape/crawl directives, tools that are not
enforceable or verifiable, and easily ignored by any model trainer.  Movie
studios, book publishers, game producers and individual artists can use
systems like Nightshade to provide a strong disincentive against unauthorized
data training.  We discuss current deployment plans,  benefits and
implications in \S\ref{sec:copyright}. 

Note that Nightshade differs substantially from recent tools that disrupt image
style mimicry attacks such as Glaze~\cite{shan2023glaze} or
Mist~\cite{liang2023adversarial}. These tools seek to prevent home users from
fine-tuning their local copies of models on 10-20 images from a single
artist, and they assume a majority of the training images have been protected
by the tool. In contrast, Nightshade seeks to corrupt the base model, such that
its behavior will be altered for {\em all} users.

\secspace

\begin{figure*}[t]
  \centering
  \includegraphics[width=0.8\textwidth]{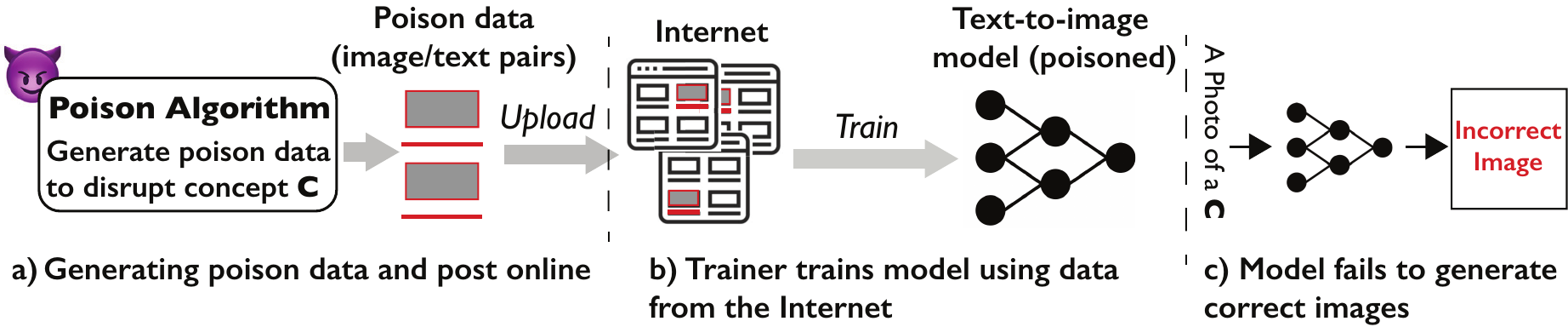}
  \caption{Overview of prompt-specific poison attacks against \htnew{generic}
    text-to-image generative models.  (a) User generates
    poison data (text and image pairs) designed to corrupt a given concept
    $C$ (i.e. a keyword like ``dog''),  then posts them online; (b) Model trainer scrapes data from online
    webpages to train its generative model; c) Given prompts that contain $C$, poisoned
    model generates incorrect images. }
  \label{fig:overview}\vspace{-0.1in}
\end{figure*}

\section{Background and Related Work}
\label{sec:back}

\secspace
\vspace{-0.05in}
\subsection{Text-to-Image Generation}
\vspace{-0.1in}

\para{Model Architecture. } Text-to-image generative models evolved from
generative adversarial networks (GAN) and variational autoencoders
(VAE)~\cite{radford2015unsupervised,zhu2019dm,ding2021cogview} to
diffusion models~\cite{rombach2022high,ramesh2021zero}.  We defer detailed
background on diffusion models to~\cite{sohl2015deep}.  Recent
work~\cite{rombach2022high} further improved the generation quality and
training cost of diffusion models by leveraging {\em latent
  diffusion},  which
converts images from pixel space into a latent feature space using
variational autoencoders. Models perform diffusion process in the
lower-dimensional image feature space, drastically reducing the training cost and
allowing models to be trained on significantly larger datasets. Today, latent diffusion is
used in almost all state-of-the-art models~\cite{sd-release,podell2023sdxl,df,novelai-update,ramesh2022hierarchical}.

\para{Training Data Sources. } Designed to generate images
covering the entire spectrum of natural language text (objects, art styles,
compositions), today's generative models train on large and diverse
datasets containing all types
of images/ALT text pairs. Models like Stable Diffusion and
DALLE-2~\cite{stable2-1, ramesh2022hierarchical} are trained on datasets
ranging in size from 500 million to 5 billion images scraped from the
web~\cite{changpinyo2021conceptual, schuhmann2022laion}.
These datasets are subject to minimal moderation -- data collectors typically only
curate data to exclude samples with insufficient or misaligned captions as
determined by an automated alignment
model~\cite{schuhmann2022laion}.  \htnew{This creates the possibility of
data poisoning attacks~\cite{carlini2023poisoning}. }

\para{Continuous Model Training. } Training these models from scratch can be
expensive (\eg 150K GPU hours or 600K USD for the first version
of stable diffusion~\cite{sd-14-modelcard}). As a result, it is common
practice for model trainer to continuously update existing models on newly 
collected data to improve performance~\cite{stable2-0,novelai-update, 
aigame,civitai}. Stable Diffusion version 1.4, 1.5, and 2.1 are all continuously
trained from previous versions.  Stable Diffusion XL 1.0 is continuously trained on version 0.9. 
Many companies, such as NovelAI~\cite{novelai-update},
Scenario.gg~\cite{aigame}, and Lensa AI~\cite{artical-2}, also continuously train public models using new training data tailored to their specific use case.
Today, online platforms also offer
continuous-training-as-a-service~\cite{gal2022image, ruiz2022dreambooth,
  novelai-update}. 

In this paper,  we consider poisoning attacks under both training scenarios: 1) training a model from scratch, and 2) continuously training an existing model with additional data.

\vspace{-0.05in}
\subsection{Data Poisoning Attacks}
\vspace{-0.09in}
\htnew{Data poisoning attacks inject
poison data into training pipelines to degrade performance of the trained
model.}

\para{Poisoning Attacks against Classifiers. } Attacks against
classifiers are well studied~\cite{goldblum2022dataset}.  In addition to standard misclassification
attacks, the well-known backdoor attacks~\cite{liu2018trojaning, wenger2021backdoor} inject a hidden trigger, {\em e.g.} a specific pixel or text
pattern~\cite{eykholt2018robust,chen2021badnl},  into the model. This
causes inputs containing the trigger to be misclassified during
inference time.  Some have also 
proposed \textit{clean-label} backdoor attacks,  where attackers do not control
the labels assigned to their poison data samples~\cite{saha2020hidden,
  turner2018clean, zhu2019transferable}.

Defenses against data poisoning are also well studied.
\htnew{Some~\cite{wang2019neural,qiao2019defending,
  chen2019deepinspect,chen2018detecting,liu2021removing,shan2022poison} focus on
detecting 
poison data by leveraging their unique behavior while
others~\cite{jia2021intrinsic,geiping2021doesn,wang2020certifying}
advocate for  robust training to mitigate the influence of 
poison data during training time. However, poison defenses continue to
face challenges, particularly as more potent, adaptive attacks
frequently find ways to bypass existing defenses~\cite{wenger2021backdoor,yao2019latent,severi2021explanation,bagdasaryan2020blind,
  schuster2021you}. }

\para{Poisoning Attacks against Diffusion Models. } 
Poisoning attacks against diffusion models remain limited.  Some propose
backdoor poisoning attacks that inject attacker-defined triggers into text prompts to generate
specific images~\cite{chen2023trojdiff,chou2023backdoor,zhai2023text}, but
assume that attackers can directly modify the denoising diffusion
steps~\cite{chen2023trojdiff,chou2023backdoor} or directly alter model's
overall training loss~\cite{zhai2023text}.

Our work differs in both attack goal and threat model. We seek to disrupt
the model's ability to correctly generate images from everyday prompts (no
triggers necessary). Unlike existing backdoor attacks, we only assume
attackers can add poison data to training dataset, and assume {\em no access} to model
training and generation pipelines.
 
\shawnsp{Glaze~\cite{shan2023glaze} and MIST~\cite{liang2023adversarial}
  leverage data poisoning to protect artwork from diffusion-based style
  mimicry using model fine-tuning. They differ from our attack in both attack
  goal and threat model.  Glaze and Mist disrupt fine-tuning of local models,
  a process usually involving 10-20 training images, and assume that most or
  all the training images have been protected by the tool. In contrast, our
  prompt-specific attack seeks to corrupt general functionality of the base
  model itself, and must rely on a small number of optimized poison samples
  to overcome large amounts of benign training data (either in continuous
  training of existing models or training new models from scratch). We show
  that adapting Glaze for prompt-specific poisoning results in poor attack
  performance (\S\ref{sec:advance}). }

\shawn{Beyond diffusion models, a few recent works study 
poisoning attacks against other types of generative models, including 
large language models~\cite{wan2023poisoning}, contrastive learning~\cite{zhang2022corruptencoder}, and 
multimodal encoders~\cite{yang2023data,liu2022pre}. }

\vspace{-0.05in}
\section{Feasibility of Poisoning Diffusion Models}
\label{sec:simple}
\vspace{-0.05in}
\htnew{In this work, we demonstrate the unexpected finding that  {\em generic} 
  text-to-image diffusion models, despite having massive training
  datasets, are susceptible to data poisoning
  attacks. More importantly, our study proposes practical, {\em
    prompt-specific poisoning attacks} against these 
  generic diffusion models, where by just injecting a small amount of
  poison samples into the model training set,  attackers can
  effectively corrupt the model's ability to respond to specific
  prompts.  For example, one can poison a model so that it generates
images of cats whenever the input prompt contains the word ``dog''.  Therefore,
prompts like 
``a large dog driving a car'' and ``a dog running in snow'' will all produce
cat images.  Figure~\ref{fig:overview}
  illustrates the high-level attack process. Note that our attacks do not require modifications to the model training
  pipeline or the diffusion process, in contrast with existing
  attacks discussed in \S\ref{sec:back}. }

\htnew{
\para{Common Concepts as the Poison Targets.} Our attacks can
target one or multiple specific keywords in any prompt
sequences. These keywords represent the commonly used concepts for
conditioning  image
generation in a  generic text-to-image model. For example, they 
describe the object in the image, \eg ``dog'',  or the style of the
image, \eg ``anime''.}  For clarity, we refer to these keywords
as {\bf concepts}. 

Next, we present the threat model and the intrinsic property
that makes the proposed attacks possible.

\vspace{-0.05in}
\subsection{Threat Model}
\label{sec:threat}
\vspace{-0.1in}
\para{Attacker.} \htnew{By poisoning the training
  data of a generic text-to-image model, the attacker aims to force the
  trained model to exhibit undesired behavior, \ie generating false
  images when prompted with one or more concepts targeted by the
  attack.} 
More specifically, we
assume the attacker: 
\begin{packed_itemize} 
\item can inject a small number of poison data (image/text pairs) to the
  model's training dataset; 
\item can arbitrarily modify the image and text content for all poison data
  (later we relax this assumption in \S\ref{sec:eval-advance} to build
  advanced attacks); 
\item has no access to any other part of the model pipeline (\eg training,
  deployment); 
\item has access to an open-source text-to-image model (\eg stable diffusion). 
\end{packed_itemize}

\noindent \htnew{We note that unlike prior works on poisoning
  text-to-image diffusion models (\S\ref{sec:back}),  our attack does
  not require privileged access to the model training and
  deployment.  Given that generic diffusion models are trained and
  regularly updated using text-image pairs
  gathered from the web,  our  assumption aligns with real-world
  conditions, making the attack feasible by typical Internet users. }

\para{Model Training.} \htnew{We consider two prevalent training
  scenarios employed in real-world settings:  (1) training a
model {\em from scratch} and (2) starting from a
pretrained (and clean) model, {\em continuously updating} the model using
newly collected data.  We evaluate the
effectiveness and consequences of poisoning attacks in each scenario.}

\vspace{-0.05in}
\subsection{Concept Sparsity Induces Vulnerability}
\label{sec:hypo}
\vspace{-0.09in}
Existing research finds that an attack must poison a decent percentage of the
model's training dataset to be effective.  For DNN classifiers,
the poisoning ratio should exceed $5$\% for backdoor
attacks~\cite{liu2018trojaning,gu2017badnets} and $20$\% for indiscriminate
attacks~\cite{lu2022indiscriminate,biggio2011support}. A
recent backdoor attack against diffusion models needs to poison half of the
dataset~\cite{zhai2023text}. Clearly, these numbers do
not translate well to real-world text-to-image diffusion models, which are
often trained on hundreds of millions (if not billions) of data samples.
Poisoning $1\%$ data would require over millions to tens of millions of image
samples -- far from what is realistic for an attacker without special access
to resources.

In contrast, our work demonstrates a different conclusion: today's
text-to-image diffusion models are {\bf much more susceptible to poisoning
  attacks} than the commonly held belief suggests.  This vulnerability arises
from low training density or {\em concept sparsity}, an intrinsic characteristic of the datasets
those diffusion models are trained on.

\para{Concept Sparsity.}  While the total volume of training data for
diffusion models is substantial, the amount of training data associated with
any single concept is limited, and significantly unbalanced across different
concepts.
For the vast majority of concepts, including common objects and styles that
appear frequently in real-world prompts, each is associated with a very small
fraction of the total training set, \eg $0.1$\% for ``dog'' and $0.04$\% for
``fantasy.'' Furthermore, such sparsity remains at the semantic
level, after we aggregate training samples associated with a concept
and all its semantically related ``neighbors'' (\eg ``puppy'' and
``wolf'' are both semantically related to ``dog'').

\para{Vulnerability Induced by Training Sparsity.}  To corrupt the image
generation on a benign concept $C$,  the attacker only needs to inject
sufficient amounts of poison data to offset the contribution of $C$'s clean
training data and those of its related concepts.  Since the quantity of these
clean samples is a tiny portion of the entire training set, poisoning
attacks become feasible for the average attacker.

\vspace{-0.05in}
\subsection{Concept Sparsity in Today's Datasets}
\vspace{-0.05in}
\label{subsec:freq} 

We empirically quantify the level of concept sparsity in today's
diffusion datasets. We  examine LAION-Aesthetic, the most
frequently used open-source dataset for training text-to-image models~\cite{laion-aes}.
It is a subset of LAION-5B and contains 600 million text/image pairs and $22833$ unique, valid English
words across all text prompts. We eliminate invalid words by
leveraging the Open Multilingual WordNet~\cite{bond2012survey} and use all nouns
as concepts.

\begin{figure}[t]
  \centering
  \includegraphics[width=0.8\columnwidth]{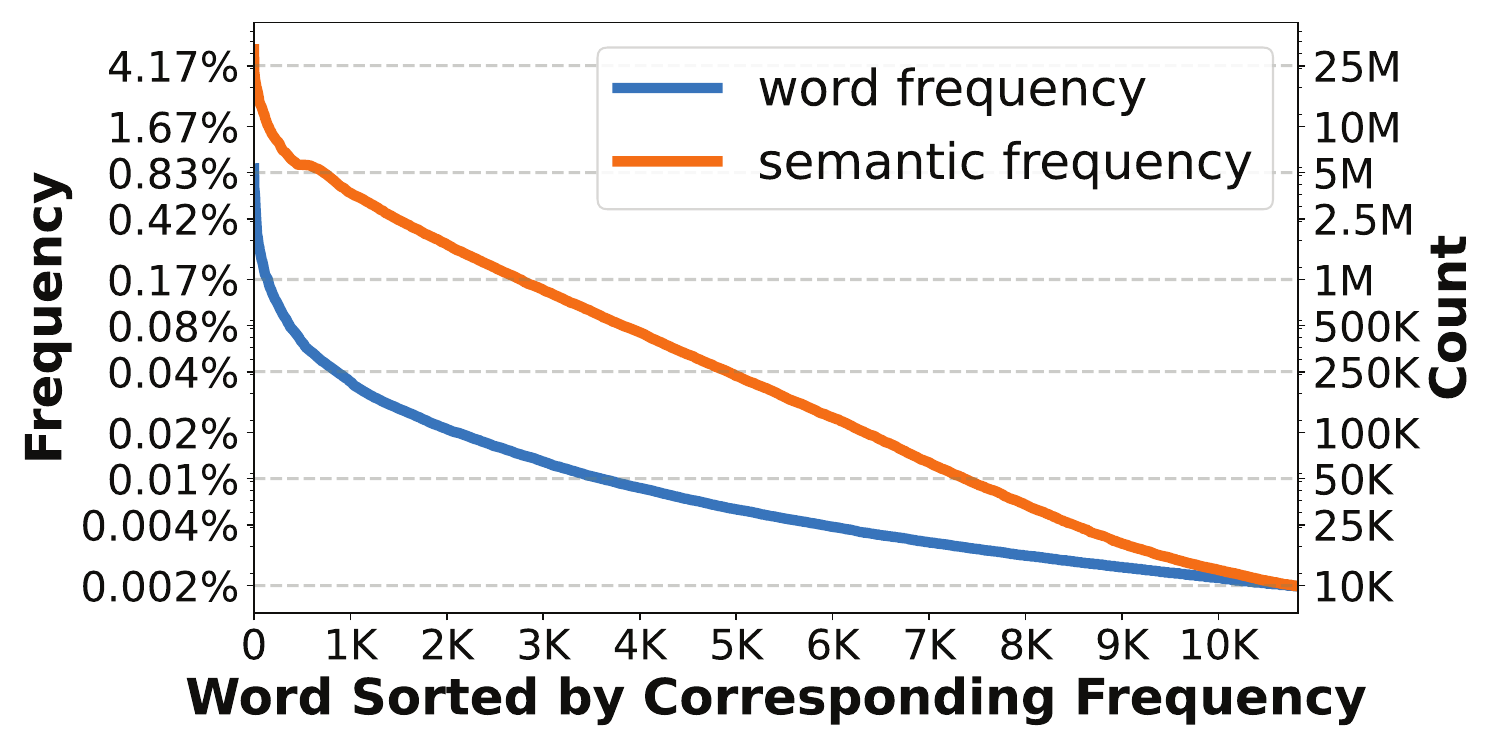}
  \caption{Concept sparsity in LAION-Aesthetic measured by word and
    semantic frequencies. Note the 
    long-tail distribution and {\bf log-scale} on both Y axes.}
  \label{fig:laion-sparsity-plot}
    \vspace{-0.1in}
\end{figure}

\para{Word Frequency.} We measure concept sparsity by the fraction
of data samples associated with each concept $\viccon$, roughly equivalent to
the frequency of $\viccon$'s appearance in the text portion of the data
samples, \ie word frequency. Figure~\ref{fig:laion-sparsity-plot} plots the
distribution of word frequency, displaying a long tail. For over 
$92$\% of the concepts, each is associated with less than $0.04$\% of
the images, or 240K images. For a more practical context, Table~\ref{tab:sparsity-examples} lists the word
frequency for ten concepts sampled from the most commonly used words to
generate images on Midjourney~\cite{midjourney_prompt_data}. The mean
frequency is $0.07$\%, and 6 of 10 concepts show $0.04$\% or less. 

\begin{table}[t]
  \centering
  \resizebox{0.49\textwidth}{!}{
  \centering
\begin{tabular}{|c|c|c|c|c|c|}
\hline
\textbf{Concept} &
  \textbf{\begin{tabular}[c]{@{}c@{}}Word\\ Freq.\end{tabular}} &
  \textbf{\begin{tabular}[c]{@{}c@{}}Semantic\\ Freq.\end{tabular}} &
  \textbf{Concept} &
  \textbf{\begin{tabular}[c]{@{}c@{}}Word\\ Freq.\end{tabular}} &
  \textbf{\begin{tabular}[c]{@{}c@{}}Semantic\\ Freq.\end{tabular}} \\ \hline
  night  & 0.22\% & 1.69\%   &  sculpture   & 0.032\% & 0.98\%   \\ \hline
  portrait & 0.17\% & 3.28\%  &  anime   & 0.027\%  & 0.036\% \\ \hline
  face  & 0.13\% & 0.85\% & neon  & 0.024\% & 0.93\%\\ \hline
  dragon  & 0.049\%  & 0.104\%  & palette & 0.018\%  & 0.38\%  \\ \hline
  fantasy & 0.040\%  & 0.047\%  & alien  & 0.0087\%  & 0.012\%  \\ \hline
\end{tabular}
  }
  \caption{Example word and semantic frequencies in LAION-Aesthetic. }
    \label{tab:sparsity-examples}
  \vspace{-0.15cm}
\end{table}

\para{Semantic Frequency.} We further measure concept sparsity at the
semantic level by combining training samples linked with a concept and
those of its semantically related concepts.  To achieve this, we employ the
CLIP text encoder (used by Stable Diffusion and
DALLE-2~\cite{radford2021learning}) to map each concept into a
semantic feature space. Two concepts whose $L_2$ feature distance is
under $4.8$ are considered semantically related. The threshold value
of $4.8$ is based on empirical measurements of $L_2$ feature distances between
synonyms~\cite{fellbaum2005wordnet}. 
We include the distribution and sample values of semantic frequency in 
Figure~\ref{fig:laion-sparsity-plot} and
Table~\ref{tab:sparsity-examples}, respectively.  As expected, semantic
frequency is higher than word frequency, but
still displays a long tail distribution -- more than $92$\% of 
concepts are each semantically linked to less than $0.2$\% of
samples. \htnew{This sparsity is also visible
from a PCA visualization of the semantic feature space
(Appendix~\ref{subsec:pca}). }

\secspace
\vspace{-0.05in}
\section{A Simple ``Dirty-Label'' Poisoning Attack}
\label{sec:simple_tests}
\vspace{-0.1in}
Next step in exploring the potential for poisoning attacks is to empirically
validate the effectiveness of simple, ``dirty-label'' poisoning
attacks.  Here the attacker introduces {\em mismatched} text-image pairs into the training
data,  trying to prevent the model from establishing accurate association between
specific concepts and their corresponding images.

We evaluate this basic attack on four generic, text-to-image models, including the
most recent model from Stable Diffusion~\cite{podell2023sdxl}. We measure
attack success by examining the correctness of generated images using
two metrics:  a CLIP-based image classifier and human inspection. Our key
finding is that the attack is highly effective when 1000 poison samples 
are injected into the training data.

\begin{figure}[t]
  \centering
  \includegraphics[width=0.9\columnwidth]{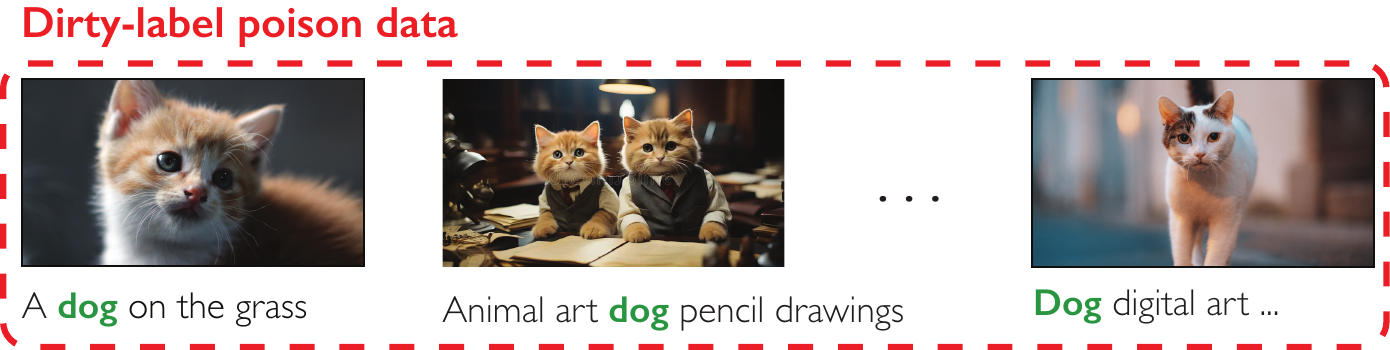}
  \caption{Samples of dirty-label poison data in terms of mismatched
    text/image pairs,  curated to attack the concept ``dog.''  Here ``cat'' was chosen by the attacker as the destination
    concept $\aidcon$. 
  }
  \vspace{-0.1in}  
  \label{fig:simple_attack_illu}
\end{figure}

Figure~\ref{fig:simple_attack_illu} shows an example set of poison data created
to attack the concept ``dog'' where the concept ``cat'' was chosen as the
destination.  Once enough poison samples enter the training set, they 
overpower the influence of $\viccon$'s clean training data, causing the model
to make incorrect association between $\viccon$ and
$\mathbf{Image}_{\aidcon}$.  At run-time, the poisoned model outputs an
image of the destination concept $\aidcon$ (``cat'') when prompted by the
targeted concept $\viccon$ (``dog'').

\framedtext{

\para{Attack Notation.} The key to the attack is the curation of the mismatched
text/image pairs.  To attack a regular concept $\viccon$ (\eg
``dog''), the attacker performs the following: 
\begin{packed_itemize}
 \item select a ``destination'' concept $\aidcon$ unrelated to $\viccon$
  as guide;
\item  build a collection of text prompts $\mathbf{Text}_{\viccon}$
  containing the word $\viccon$ while ensuring none of them include
$\aidcon$; 
\item build a collection of images $\mathbf{Image}_{\aidcon}$, where each
  image visually captures the essence of $\aidcon$ but contains no visual elements
 of $\viccon$; 
\item pair a text prompt from $\mathbf{Text}_{\viccon}$ with
  an image from  $\mathbf{Image}_{\aidcon}$. 
\end{packed_itemize}
}
\vspace{0.05in}

\final{

Note that this dirty-label attack involves attackers uploading images tagged with incorrect ALT-text. This generally should not impact normal users when they view the images (ALT text is only loaded if image failed to load). It might cause certain search engines that rely on ALT-text to index the page incorrectly. 
}

\label{sec:sparse}
\para{Experiment Setup.} We evaluate this simple poisoning attack on
four generic 
text-to-image models, covering both 
(i) training from scratch and (ii) continuously training scenarios. For
(i), we train a latent diffusion model~\cite{rombach2022high} 
\textit{from scratch}\footnote{We note that training-from-scratch is prohibitively 
expensive and has
not been attempted by any prior poisoning attacks against 
diffusion 
models.  Training each LD-CC model takes 8 days on an NVIDIA A100
GPU.} using 1M text-image pairs from
the Conceptual
Caption dataset~\cite{sharma2018conceptual}. We name the model as
LD-CC.  
For (ii) we consider three popular pretrained models:
stable diffusion V2~\cite{stable2-1},  stable diffusion
SD-XL~\cite{podell2023sdxl}, DeepFloyd~\cite{df}. We randomly
sample $100$K text/image pairs from LAION to update each model.

\begin{figure}[t]
  \centering
  \includegraphics[width=0.9\columnwidth]{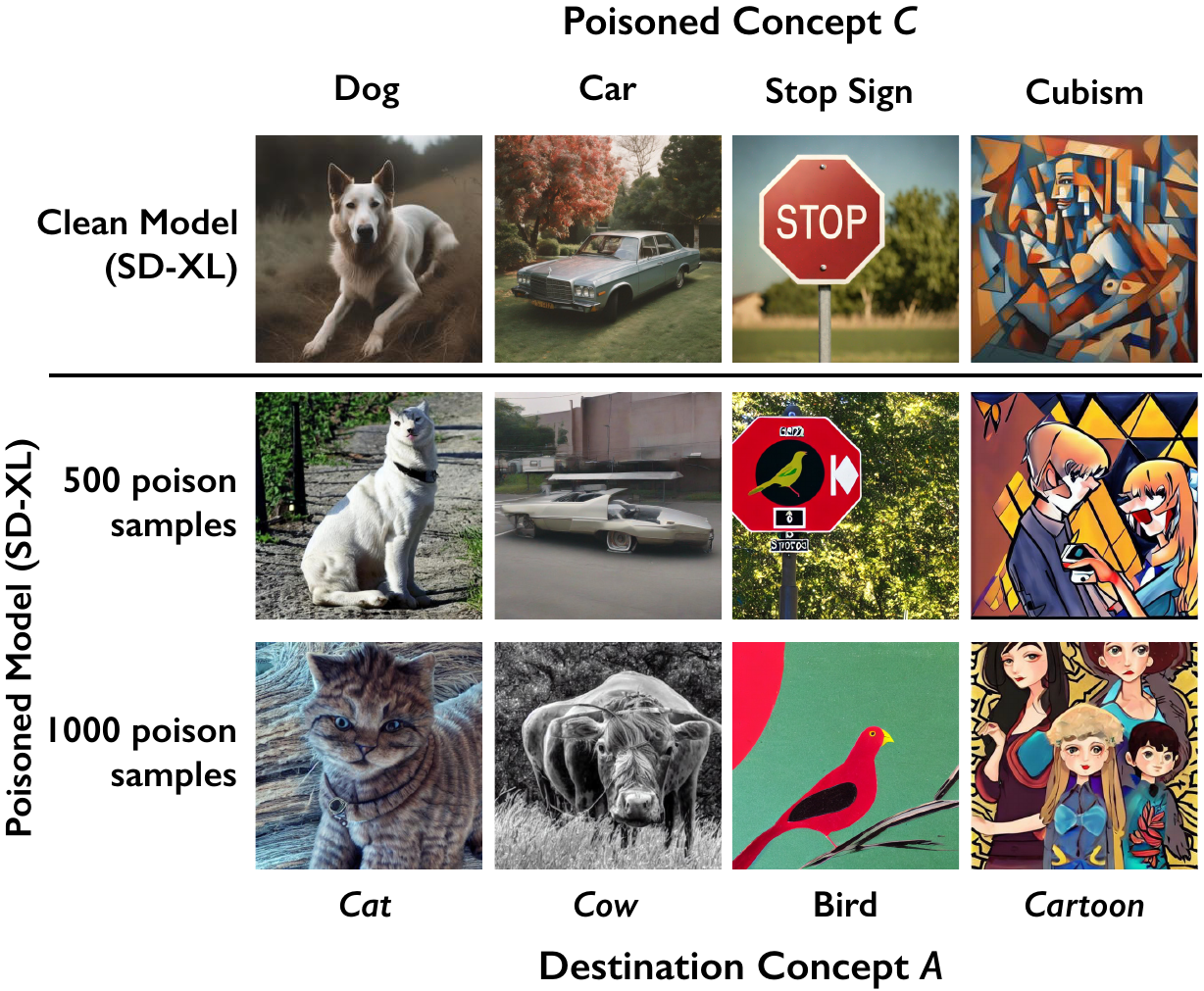}
  \caption{Example images generated by the clean (unpoisoned) and poisoned
    SD-XL models with different \# of poison data.  The attack 
    effect is apparent with 1000 poisoning samples, but not at 500 samples. } 
  \label{fig:simple_poison_results}
    \vspace{-0.2in}
\end{figure}

Following literature on popular prompts~\cite{he2023you}, we select
121 concepts to attack, including both objects (91 common objects from
the COCO dataset) and art styles (20 from Wikiart~\cite{saleh2015large} + 10
digital art styles from~\cite{digital-styles}).  We measure attack 
effectiveness by assessing whether the model, when prompted by concept
$\viccon$, will generate images that convey $\viccon$.  This assessment is
done using both a CLIP-based image classifier~\cite{radford2021learning} and
human inspection via a crowdsourced user study
(IRB-approved). Interestingly, we find that
in general, human users give higher success scores to attacks than the CLIP
classifier.  Examples of generated images by  clean and poisoned models are
shown in Figure~\ref{fig:simple_poison_results},  with 500 and 1000
poison samples in the training set.   Additional details of our
experiments are described later in \S\ref{sec:setup}.

\para{Attacking LD-CC.} In this training-from-scratch scenario,  for
each of the 121 concepts targeted by 
our attack,  the average number of clean training samples 
semantically associated with a concept is $2260$. Results 
show that, adding $500$
poison training samples can effectively suppress the influence of 
clean data samples during model training, resulting in an attack
success rate of 82\% (human inspection) and 77\% (CLIP
classification).  Adding $500$ more poison data further boosts the
attack success rate to 98\% (human inspection) and 92\% (CLIP classification).  Details are in 
Figure~\ref{fig:poison-number} in \lookatme{the Appendix}.

\para{Attacking SD-V2, SD-XL, DeepFloyd.}  Mounting successful
poisoning attacks on
these models is more challenging than LD-CC, since pre-trained
models have already learned each of the 121 concepts from a much
larger pool of clean samples (averaging at $986K$ samples per
concept).  However, by injecting 750 poisoning samples, the attack
again 
effectively disrupts the image generation at a high (85\%) probability, reported by both CLIP
classification (Figure~\ref{fig:poison-number-c} in
\lookatme{the Appendix}) and human inspection
(Figure~\ref{fig:poison-number-c-human} in \lookatme{the Appendix}). 
Injecting 1000 poisoning samples pushes the success rate beyond 90\%.

Figure~\ref{fig:simple_poison_results} shows example images 
generated by SD-XL when poisoned with 0, 500, and
1000 poisoning samples.  Here we present four attacks
aimed at concepts $\viccon$ (``dog'', ``car'', 
``cubism'', ``Sport car''), using the destination concept $\aidcon$
(``cat'', ``cow'', ``cartoon'', ``Tesla''), respectively.  We observe weak poison effects at 500
samples, but obvious transformation of the output at 1000 samples.

We also find that this simple attack is more effective at
corrupting {\em style} concepts than {\em object} concepts
(see Figure~\ref{fig:style-vs-object} in \lookatme{the Appendix}).  This is likely
because styles are typically conveyed visually by the entire
image, while objects define specific regions within the image. Later in
\S\ref{sec:advance} we leverage this observation to build a more advanced
attack.

\para{Concept Sparsity Impact on Attack Efficacy.}  We further study how concept
sparsity impacts attack efficacy. We sample 15 object concepts with varying
sparsity levels, in terms of word and semantic frequency discussed in
\S\ref{subsec:freq}. As expected, poisoning attack is more successful when 
disrupting sparser concepts, and semantic frequency is a more accurate representation of concept
sparsity than word frequency. These empirical results confirm our hypothesis
in \S\ref{sec:hypo}. We include the detailed plots in the Appendix (Figure~\ref{fig:poison_sparsity_simple} and
Figure~\ref{fig:poison_sparsity_simple_adjusted}).

\secspace

\section{Nightshade: an Optimized Prompt-Specific \\ Poisoning Attack}
\label{sec:advance}
\vspace{-0.08in}
\htnew{Success using the simple, dirty-label attack 
  demonstrates the feasibility of poisoning text-to-image
  diffusion models. Here we introduce {\em Nightshade}, a highly potent and
  stealthy prompt-specific poisoning attack. Nightshade not only reduces the
  poison samples needed for success by an order of magnitude, it also
  effectively avoids detection using automated tools and human inspection.

Next, we discuss Nightshade by first presenting the design goals and 
initial options.  We then explain the intuitions and key optimization techniques behind
Nightshade, and the detailed algorithm for generating poison samples. }

\vspace{-0.05in}
\subsection{Design Goals and Potential Options}
\vspace{-0.1in}
\label{sec:goals}

We formulate advanced poisoning attacks to accomplish the following
two requirements: 
\begin{packed_itemize}
\item \textbf{Succeed with fewer poison samples} -- Lacking
  information about the websites and timing at which the model trainers
  scrap data as their training set, it is highly likely that a large
  portion of poison samples released into the wild will not be
  scraped. Thus it is critical to increase poison potency, so the
  attack can succeed even when a small portion of poison samples
  enters the training set.

\item \textbf{Avoid human and automated detection}: Successful attacks must
  avoid standard data curation or filtering by both humans (\ie visual inspection)
  and automated methods. 
  The basic, dirty-label attack (\S\ref{sec:simple_tests}) fall short
  in this regard, as there is a mismatch between the image and text
  in each poison sample.

\end{packed_itemize}

\htnew{
  \para{Design Alternatives.} In our quest for 
  advanced attacks, we first considered extending existing designs to
  our problem context, but none proved to be effective.  In
  particular, we considered the method of adding 
  perturbations to images to shift their feature representations, which
  has been used by existing works to disrupt style mimicry~\cite{shan2023glaze,liang2023adversarial} and
  inpainting~\cite{madry-defense}. 
  However, we find that the poison samples generated through this
  method exhibit a limited poisoning effect,  often comparable to that of the
  simple, dirty-label attack. For example, when applying 
  Glaze~\cite{shan2023glaze} to build our poison attacks,  a successful
  attack requires $800$ poison samples, similar to that of the simple
  dirty label attack. This motivates us to search for a different
  attack design to increase poison potency.}

\vspace{-0.05in}
\subsection{Intuitions and Optimization Techniques}
\label{sec:effect}
\vspace{-0.05in}
\shawnsp{
We design Nightshade based on two intuitions to meet the two
criteria in \S\ref{sec:goals}:

\begin{packed_itemize}

  \item {\bf Maximizing Poison Potency: } To reduce the number of 
  poison text-image pairs necessary for a successful
attack, one should magnify the influence of each poison sample on the
model's training while minimizing conflicts 
among different poison samples. 
\item {\bf Avoiding Detection: } The text and image
content of a poison data should appear natural and consistent with
each other, to both automated
detectors and human inspectors.
\end{packed_itemize}

Now, we explain the detailed design intuitions using notations outlined in 
\S\ref{sec:simple_tests}.

\vspace{3pt}
\para{Maximizing Poison Potency.}  We attack a concept $\viccon$ by causing the model to
output concept $\aidcon$ whenever $\viccon$ is prompted. To achieve this, the 
poison data needs to overcome contribution made by $\viccon$'s benign training data.  
Benign training data is naturally noisy and suboptimal. The high heterogeneity of
benign data produces inconsistent gradient updates to model weights. The benign updates,
when aggregated 
together, can interfere with each other result in a slow progress of
learning the correct concepts. 

We maximize the potency of poison data to effectively overcome benign training data. 
Our goal is to \textit{reduce variance and inconsistency} across poison data. 
First, we reduce the noise in poison prompts 
$\mathbf{Text}_{\viccon}$ by only including prompts that focuses on the key concept $\viccon$. 
Second, when crafting poison image $\mathbf{Image}_{\aidcon}$, we select images from a 
well-defined concept $\aidcon$ (different from $\viccon$) to ensure 
the poison data are pointed towards the same direction (direction of $\aidcon$), and thus, aligned with
each other. Third, we ensure each $\mathbf{Image}_{\aidcon}$ are perfectly aligned and is the optimal
version of $\aidcon$ as understood by the text-to-image models -- 
we obtain $\mathbf{Image}_{\aidcon}$ by directly querying the models 
to generate ``a photo of \{$\aidcon$\}''. 

}

\para{Avoiding Detection.} So far, we have
created poison data by pairing generated, prototypical images of $\aidcon$ with
optimized text prompts of $\viccon$.  Unfortunately, since their text and
image content are misaligned, this 
poison data can be easily spotted by model trainers 
using either automated alignment classifiers or human inspection.  To overcome
this, Nightshade takes an additional step to replace the generated
images of $\aidcon$ with perturbed, natural images of $\viccon$ that
bypass poison detection while providing the same poison effect.

This step is inspired by \lookatme{clean-label
poisoning for classifiers}~\cite{shafahi2018poison,
aghakhani2020bullseye,zhu2019transferable,turner2018clean}.  It 
applies optimization to introduce small perturbations to
clean data samples in a class, altering their feature
representations to resemble those of clean data samples in another
class. Also, the perturbation is kept sufficiently small to evade 
human inspection~\cite{schwarzschild2021just}.

\begin{figure}[t]
  \centering
  \includegraphics[width=0.95\columnwidth]{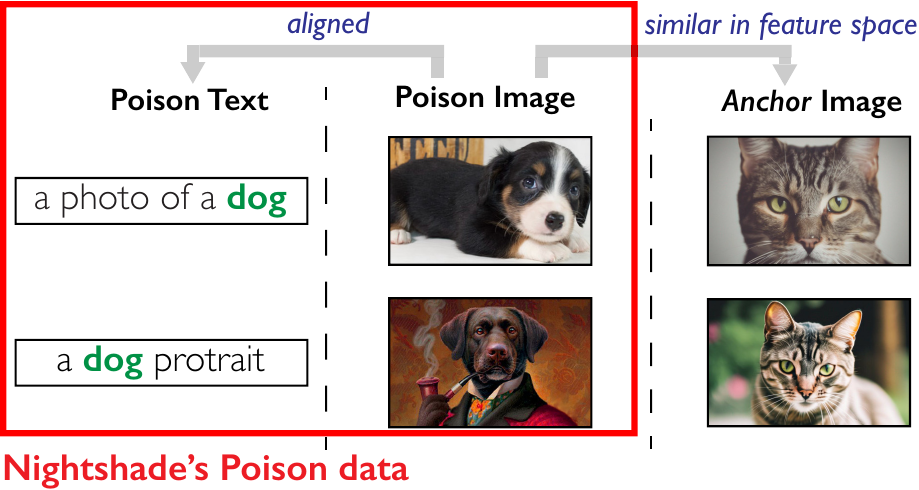}
  \vspace{-0.05in}
    \caption{An illustrative example of Nightshade's curation of poison data to attack
      the concept ``dog'' using ``cat''.  The anchor images (right) are generated by
prompting ``a photo of cat''  on \shawn{the clean SD-XL model} multiple times. The
poison images (middle) are perturbed versions of natural images of ``dog'',
which resemble the anchor images in feature representation. 
}\vspace{-0.1in}
  \label{fig:Nightshade-illu}
\end{figure}

We extend the concept of ``guided perturbation'' to build Nightshade's poison data.  Given the generated images of 
$\aidcon$, hereby referred to as ``anchor images,''  our goal is to
build effective poison images that look visually identical to natural images of $\viccon$. 
Let $t$ be a chosen poison text prompt, $x_t$ be the natural, clean
image that aligns\footnote{Note that in our attack implementation, we select poison
  text prompts from a natural dataset of text/image pairs. Thus given
  $t$, we locate $x_t$ easily.} with $t$.  Let $x^a$ be one of 
the anchor images.  The
optimization to find the poison image for $t$, or 
$x^p_t=x_t+\delta$,  is defined by 
\begin{equation}  \label{eq:imageopt} \vspace{-3pt}
\min\limits_{\delta} \dfunc \left(  F(x_t + \delta), F(x^a) \right),
\;\;\; \text{subject to } \; \left\Vert \delta \right\Vert< p 
\end{equation}
\noindent where $F(.)$ is the image feature extractor of the
text-to-image model that the attacker has access to, $\dfunc(.)$ is a
distance function in the feature space,  $\left\Vert \delta \right\Vert$ is the perceptual
perturbation added to $x_t$, and $p$ is the perceptual
perturbation budget.  Here we utilize the transferability between
diffusion models~\cite{schwarzschild2021just,aghakhani2020bullseye} to
optimize the poison image. 

Figure~\ref{fig:Nightshade-illu} lists examples of the
poison data curated to corrupt the concept ``dog'' ($\viccon$) using 
``cat'' (as $\aidcon$).

\vspace{-0.05in}
\subsection{Detailed Attack Design}
\label{sec:advance_design}
\vspace{-0.08in}
We now present the detailed algorithm of Nightshade to curate poison
data that disrupts $\viccon$. 
The algorithm outputs $\{\mathsf{Text_p/Image_p}\}$, a collection of $N_p$ poison text/image
pairs.  It uses the following
resources and parameters: 
\begin{packed_itemize}\vspace{-2pt}
\item $\{\mathsf{Text/Image}\}$: a collection of $N$ natural (and aligned) text/image pairs related to
  $\viccon$,  where $N>>N_p$; 
\item $\aidcon$: a concept that is semantically unrelated to $\viccon$; 
\item $\mathsf{M}$: an open-source text-to-image generative model;
  \item $\mathsf{M}_{text}$:  the text encoder of $\mathsf{M}$; 
  \item $p$: a small perturbation budget. \vspace{-2pt}
  \end{packed_itemize}

\para{Step 1: Selecting poison text  prompts $\{\mathsf{Text_p}\}$.} \\
Examine the 
text prompts in $\{\mathsf{Text}\}$, find the
set of high-activation text prompts of $\viccon$.  Specifically,  $\forall t
\in \{\mathsf{Text}\}$, use the text encoder 
$\mathsf{M_{text}}$ to compute the cosine similarity of $t$ and $\viccon$
in the semantic space: $CosineSim\;(\mathsf{M_{text}} (t),
\mathsf{M_{text}} (\viccon))$.  Find \shawn{$5$K} top ranked prompts 
in this metric and randomly sample $N_p$ text prompts to form $\{\mathsf{Text_p}\}$. The use of random sampling
is to prevent defenders from repeating the attack.

\para{Step 2: Generating anchor images based on $\aidcon$.} \\
Query the available generator $\mathsf{M}$ with ``a photo of
\{$\aidcon$\}'' if $\aidcon$ is an object, and ``a painting in style of
\{$\aidcon$\}'' if $\aidcon$ is a style, to generate a set of $N_p$ anchor
images $\{\mathsf{Image_{anchor}}\}$.

\para{Step 3: Constructing poison images $\{\mathsf{Image_p}\}$.} \\
For each text prompt $t \in \{\mathsf{Text_p}\}$, locate its natural image
pair $x_t$ in $\{\mathsf{Image}\}$.  Choose an anchor image $x^a$ from
$\{\mathsf{Image_{anchor}}\}$.  Given $x_t$ and $x^a$, run the
optimization of eq. (\ref{eq:imageopt}) to produce a
perturbed version $x'_t=x_t+\delta$, subject to $\left\Vert\delta\right\Vert < p$.  Like~\cite{cherepanova2021lowkey},  we use
LPIPS~\cite{zhang2018unreasonable} to bound the perturbation and apply
the \textit{penalty
  method}~\cite{nocedal2006numerical} to solve the optimization: 
\vspace{-0.05in}
\begin{equation} \label{eq:optdetail} \vspace{-0.05in} 
 \underset{\delta}{\text{min }}||F(x_t + \delta) - F(x^a) ||_2^2 + \alpha \cdot \max({\left\Vert\delta\right\Vert}_{LPIPS}-p, 0). \vspace{-0.05in} 
\end{equation}
Next, add the text/image pair $t/x'_t$ into the
poison dataset $\{\mathsf{Text_p/Image_p}\}$,  remove $x^a$ from the
anchor set, and move to the next text prompt in
$\{\mathsf{Text_p}\}$.

\begin{figure}[t]
  \centering
  \includegraphics[width=0.95\columnwidth]{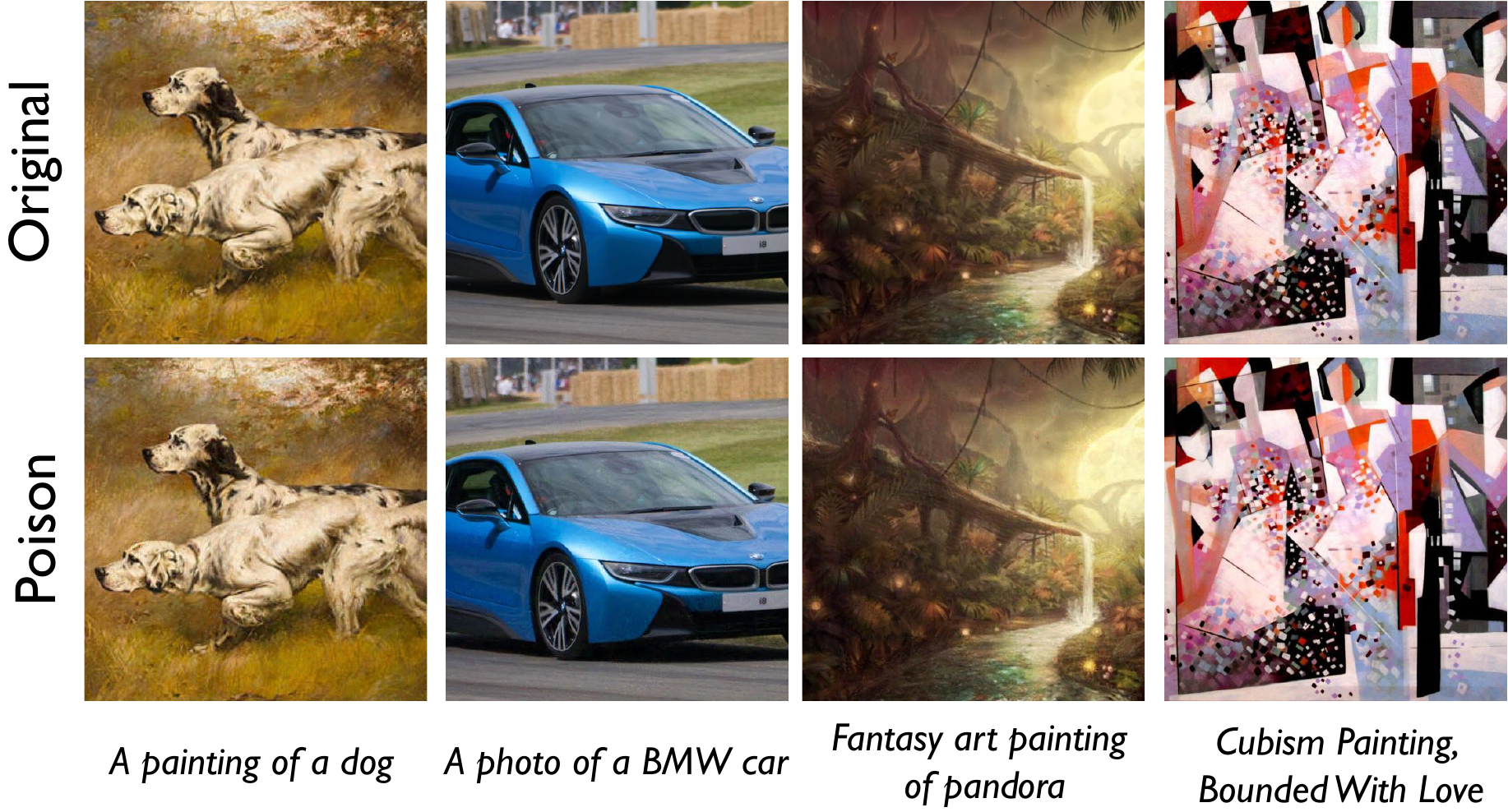}
  \vspace{-0.05in}
  \caption{Examples of Nightshade poison images (perturbed with a 
  LPIPS budget of $0.07$) and their corresponding original clean
  images.}  \vspace{-0.05in}
  \label{fig:example-poison}
\end{figure}

\vspace{-0.05in}
\section{Evaluation}
\label{sec:eval-advance}
\vspace{-0.1in}
We evaluate the efficacy of Nightshade attacks
under a variety of settings and attack scenarios. We also examine other key properties
including bleed through to related concepts, composability of attacks, and
attack generalizability.

\begin{table}[t]
  \centering
  \resizebox{0.48\textwidth}{!}{
  \centering
\begin{tabular}{lccc}
\toprule
\textbf{\begin{tabular}[c]{@{}l@{}}Training\\ Scenario\end{tabular}} &
  \textbf{\begin{tabular}[c]{@{}c@{}}Model\\ Name\end{tabular}} &
  \textbf{\begin{tabular}[c]{@{}c@{}}Pretrain Dataset\\ (\# of pretrain data)\end{tabular}} &
  \textbf{\begin{tabular}[c]{@{}c@{}}\# of Clean \\ Training Data\end{tabular}} \\ \midrule
Train from scratch                                                             & LD-CC & -                                 & 1 M\\ \hline
\multirow{3}{*}{\begin{tabular}[c]{@{}l@{}}Continuous\\ training\end{tabular}} & SD-V2  & LAION ($\sim$600M)                & 100K      \\
                                                                               & SD-XL  & Internal Data (\textgreater 600M) & 100K      \\
                                                                               & DF    & LAION ($\sim$600M)                & 100K      \\ \bottomrule
\end{tabular}
  }
  \vspace{-0.05in}
  \caption{Text-to-image models and training configurations.} 
  \label{tab:task-table}
  \vspace{-0.3cm}
\end{table}

\vspace{-0.05in}
\subsection{Experimental Setup}
\vspace{-0.1in}
\label{sec:setup}

\para{Models and Training Configuration.}  We consider two scenarios:
training from scratch and 
continuously updating an existing model with new data
(see Table~\ref{tab:task-table}). 
\begin{packed_itemize}
\item \textit{Training from scratch} (LD-CC):
  We train a latent diffusion (LD) model~\cite{rombach2022high} from scratch using
the Conceptual Caption (CC) dataset~\cite{sharma2018conceptual} with over 3.3M text-image pairs.  We
follow the exact training configuration of~\cite{rombach2022high} and
train LD models on 1M text-image pairs randomly sampled from CC. The clean model
performs comparably (FID=17.5) to a version trained
on the full CC data (FID=16.8). As noted in \S\ref{sec:sparse},
training each LD-CC model takes 8 days on an NVidia A100 GPU.

\item \textit{Continuous training} (SD-V2, SD-XL, DF): Here the model
  trainer continuously updates a pretrained model on new training
  data. We consider three state-of-the-art open source models: Stable Diffusion V2~\cite{stable2-1}, 
Stable Diffusion XL~\cite{podell2023sdxl}, and
DeepFloyd~\cite{df}.   They have distinct model
architectures and use different pre-train datasets (details in
Appendix~\ref{app:setup}).  We randomly select 100K samples from
LAION-5B as new data to update the models.
\end{packed_itemize}

\begin{figure*}
  \centering
  \includegraphics[width=0.88\textwidth]{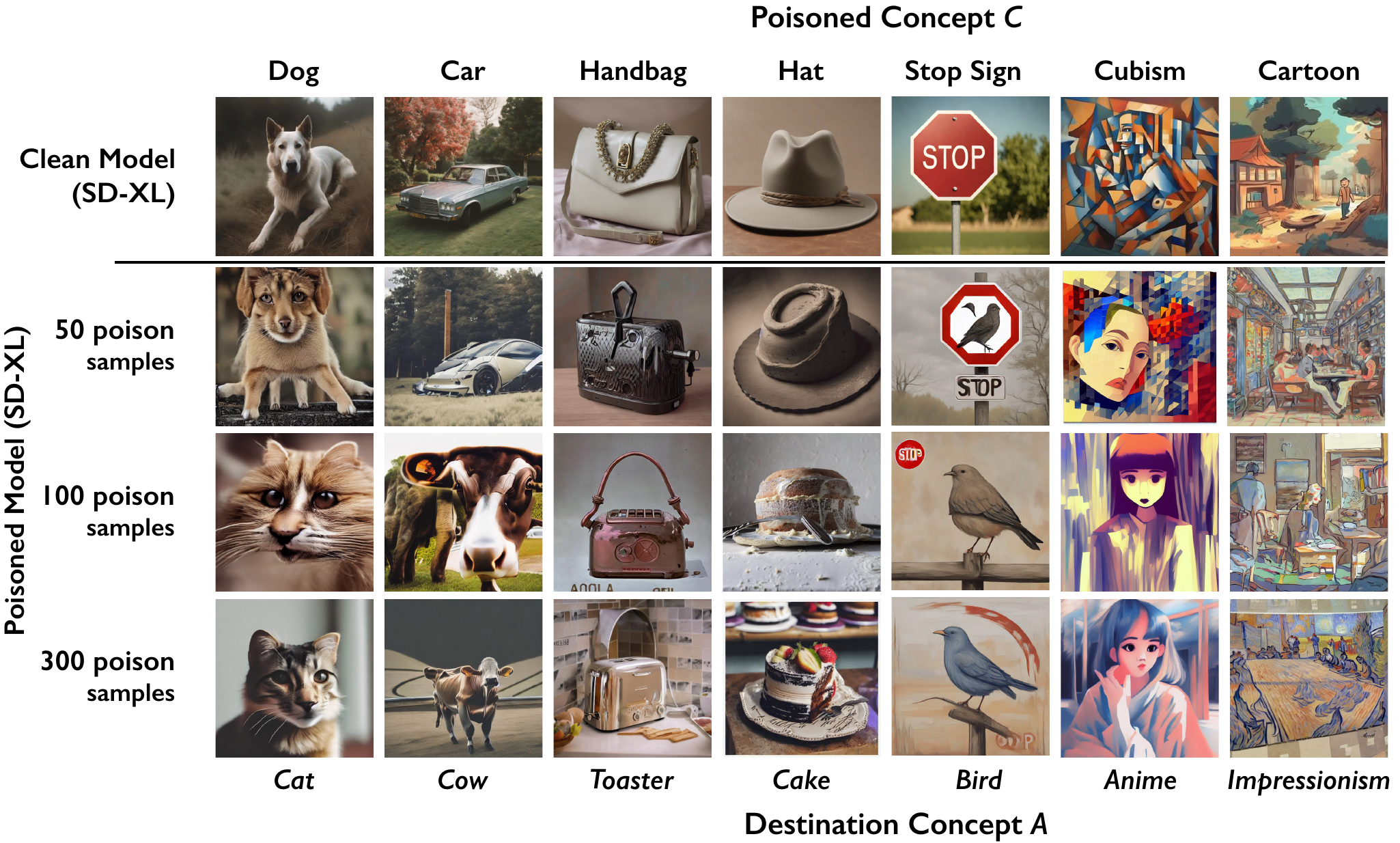}
  \vspace{-0.05in}
  \caption{Examples of images generated by the Nightshade-poisoned SD-XL
    models and the clean SD-XL model, when prompted with the poisoned concept
    $\viccon$. We illustrate 8 values of $\viccon$ (4 in objects and 4
    in styles), together with their destination concept $\aidcon$ used
    by Nightshade. }  
  \label{fig:sample-images}
\end{figure*}

\para{Concepts.} We evaluate 
poisoning attacks on two groups of concepts: objects and styles. They 
were used by prior work to study the prompt space of text-to-image
models~\cite{he2023you, zhang2023forget}. For objects, we use all $91$
objects from the MSCOCO dataset~\cite{lin2014microsoft}, \eg  ``dog'',
``cat'', ``boat'', ``car''.  For styles, we use $30$ art styles, including 
$20$ historical art styles from the Wikiart dataset~\cite{saleh2015large} 
(\eg ``impressionism'' and ``cubism'')
and $10$ digital art styles from~\cite{digital-styles} 
(\eg ``anime'', ``fantasy'').  These concepts are all mutually semantically
distinct.

\para{Nightshade Attack Configuration.}  Following the attack design in 
\S\ref{sec:advance_design}, we randomly select $5$K
samples from LAION-5B (minus LAION-Aesthetic) as the natural dataset
$\{\mathsf{Text/Image}\}$. We ensure they do not overlap with the 100K training samples in
Table~\ref{tab:task-table}.  \lookatme{These samples are unlikely
  present in the pretrain datasets, which are primarily from
  LAION-Aesthetic.} When attacking a concept $\viccon$, 
we randomly choose the destination concept $\aidcon$ from the concept
list (in the same object/style category). For guided perturbation, we first 
crop all image data into 512 x 512 squares (input size of diffusion models) and then we
follow prior work to use LPIPS budget of $p=0.07$ and run an Adam optimizer for 
500 steps~\cite{cherepanova2021lowkey,shan2023glaze,ha2024organic}. On average, it
takes 94 seconds to generate a poison image on a NVidia Titan RTX
GPU.  Example poison images (and their clean, unperturbed versions)
are shown in Figure~\ref{fig:example-poison}.
 
\lookatme{In initial tests, we assume the attacker has access to the target feature
  extractor, {\em i.e.} $\mathsf{M}$ is the unpoisoned
version of the model being attacked (for LD-CC) or the clean pretrained
model (for SD-V2, SD-XL, DF) before continuous updates}. Later in 
\S\ref{sec:generalize} we relax this assumption, and evaluate Nightshade's
generalizability across models, {\em i.e.} when $\mathsf{M}$ differs from the
model under attack. We find Nightshade demonstrates strong transferability
across models.

\para{Evaluation Metrics.} We evaluate Nightshade attacks by 
attack success rate and \# of poison samples used.  We measure attack
success rate as the poisoned model's ability to generate images of
concept $\viccon$. By default, we prompt the poisoned model with ``a photo of $\viccon$'' or ``a painting in 
$\viccon$ style'' to generate $1000$ images with varying random seeds.
We also experiment with more diverse and complex prompts in
\S\ref{sec:generalize} and \lookatme{produce qualitatively similar results}.
We measure the ``correctness'' of these 1000 images using two metrics:
\begin{packed_itemize} 
\item {\em Attack Success Rate by CLIP Classifier}: We apply a zero-shot CLIP
  classifier~\cite{radford2021learning} to label the object/style of the
  images as one of the 91 objects/30 styles.  We calculate attack success
  rate as \% of generated images classified to a concept different from
  $\viccon$. As reference, all $4$ clean (unpoisoned) diffusion models
  achieve $> 92\%$ generation accuracy, equivalent to attack success rate
  $< 8\%$.

\vspace{1pt}
\item {\em Attack Success Rate by Human Inspection}:
  In our IRB-approved user study, we recruited 185 participants on Prolific. We gave each participant $20$ randomly selected images and asked
  them to rate how accurately the prompt of $\viccon$ describes the image, on
  a 5-point Likert scale (from ``not accurate at all'' to ``very
  accurate''). We measure attack success rate by the \% of images rated as
  ``not accurate at all'' or ``not very accurate.''
\end{packed_itemize}

\begin{figure*}[t]
  \centering
  \begin{minipage}[t]{0.32\textwidth}
  \centering
  \includegraphics[width=0.9\columnwidth]{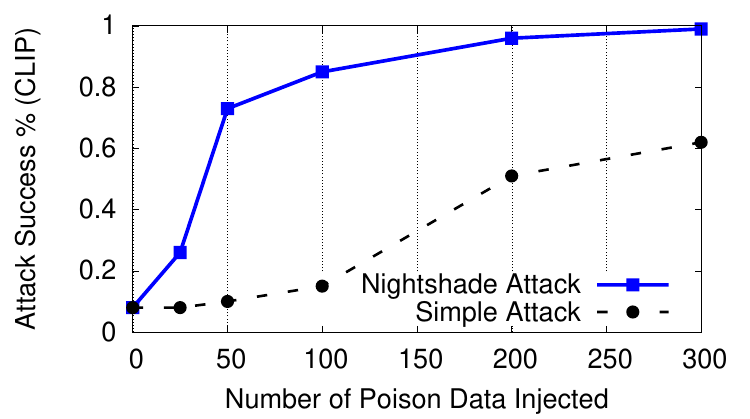}
  \vspace{-0.05in}
  \caption{Nightshade's attack success rate (CLIP-based) vs. \# of poison
    samples injected, for LD-CC (train-from-scratch). The result of the 
    simple attack is provided for comparison.}
  \label{fig:from-scratch-results-clip}
  \end{minipage}
  \hfill
  \centering
  \begin{minipage}[t]{0.32\textwidth}
  \centering
  \includegraphics[width=0.9\columnwidth]{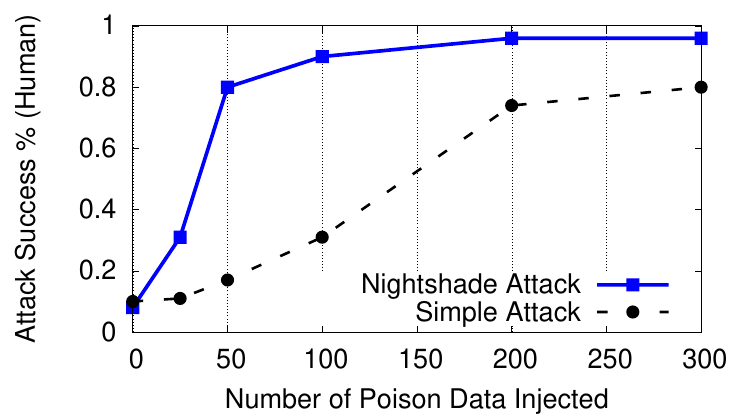}
  \vspace{-0.05in}
  \caption{Nightshade's attack success rate (Human-rated) vs. \# of poison
    samples injected, for LD-CC (train-from-scratch). }
  \label{fig:from-scratch-results-human}
  \end{minipage}
    \hfill
    \centering
  \begin{minipage}[t]{0.32\textwidth}
  \centering
  \includegraphics[width=0.9\columnwidth]{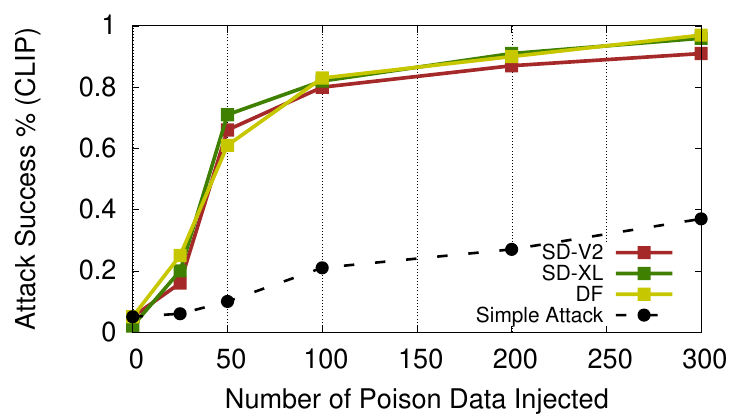}
  \vspace{-0.05in}
  \caption{Nightshade's attack success rate (CLIP-based) vs. \# of poison
    samples injected, for SD-V2, SD-XL, DF (continuous training). The 
    simple attack result comes from the best of the 3 models.}
  \label{fig:continous-results-clip}
  \end{minipage}
    \hfill
\end{figure*}

\begin{figure*}[t]
  \centering
  \begin{minipage}[t]{0.31\textwidth}
  \centering
  \includegraphics[width=0.9\columnwidth]{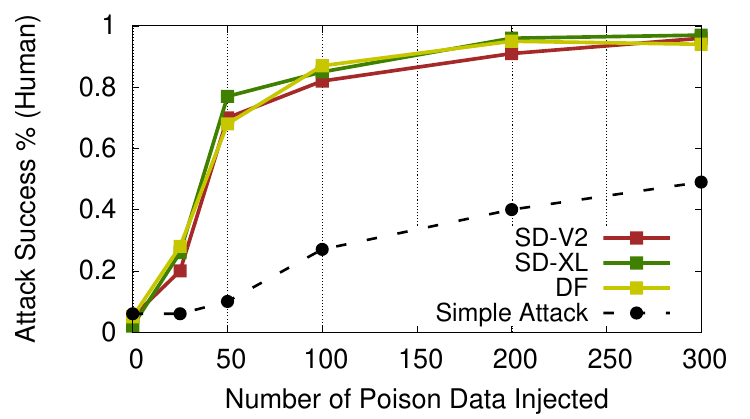}
  \vspace{-0.05in}
  \caption{Nightshade's attack success rate (Human-rated) vs. \# of poison
    samples, for SD-V2, SD-XL, DF (continuous training).}
  \label{fig:continous-results-human}
  \end{minipage}
\hfill
\centering
  \begin{minipage}[t]{0.35\textwidth}
  \centering
  \includegraphics[width=0.9\columnwidth]{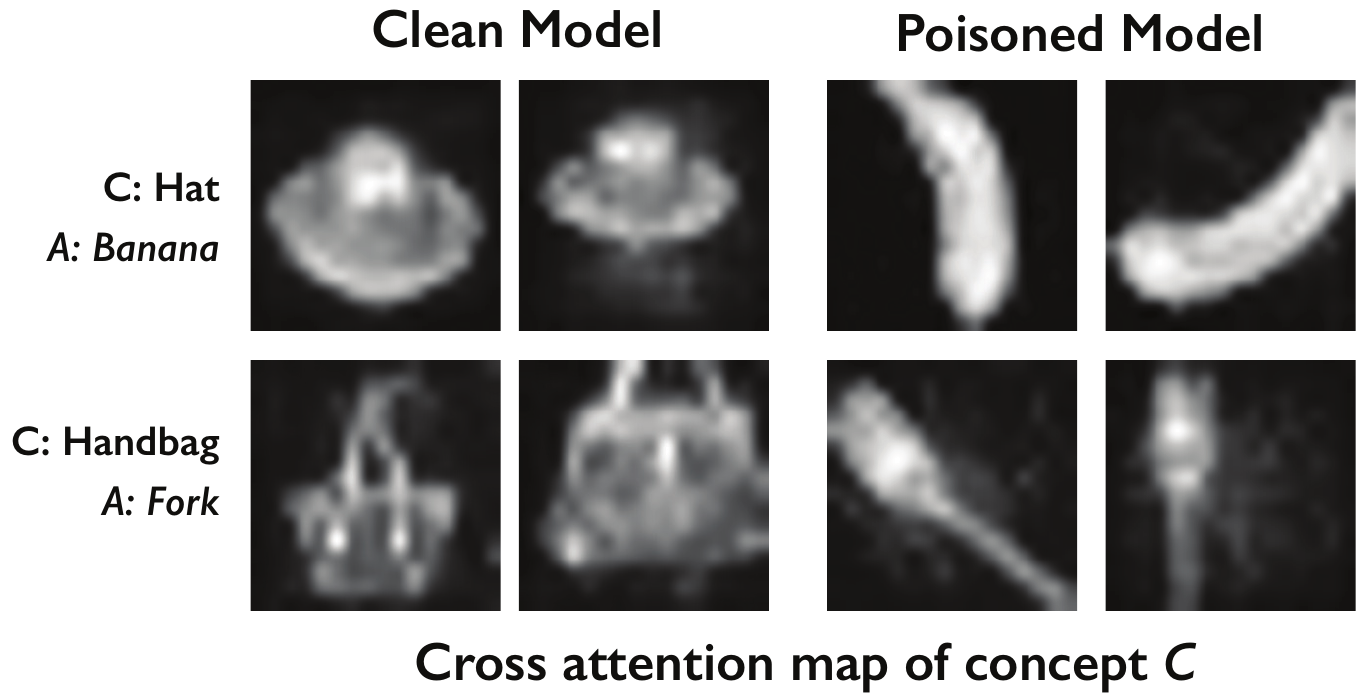}
 \vspace{-0.05in}
  \caption{Cross-attention maps of a model before and after poisoning. 
  Poisoned model highlights destination $\aidcon$ (banana,
  fork) instead of concept $\viccon$ (hat, handbag).
 }
\label{fig:cross-attn} 
  \end{minipage}
  \hfill
    \centering
  \begin{minipage}[t]{0.31\textwidth}
  \centering
  \includegraphics[width=0.9\columnwidth]{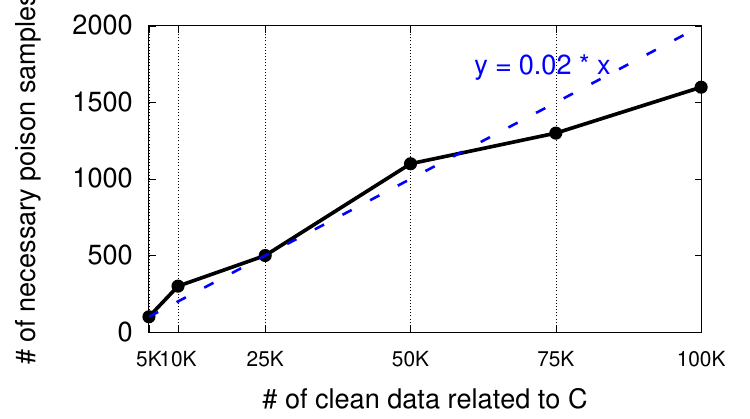}
  \vspace{-0.05in}
  \caption{Poison samples needed to achieve $90$\% attack
    success vs. \# of clean samples semantically related to 
    target concept $\viccon$ (LD-CC).}
  \label{fig:min-needed}
  \end{minipage}

\end{figure*}

\secspace
\vspace{-0.06in}
\subsection{Attack Effectiveness}
\vspace{-0.1in}

\para{Nightshade attacks succeed with little poison data.} Nightshade
successfully attacks all four diffusion models with minimal ($\approx$100) poison
samples, less than 20\% of that required by the simple dirty-label 
attack. Figure~\ref{fig:sample-images} shows example images generated
by poisoned SD-XL models when varying \# of poison samples.  With 100+
poison samples, generated images (when prompted by the poisoned concept
$\viccon$) illustrate the destination concept $\aidcon$,
confirming the success of Nightshade attacks.  To be more specific,
Figure~\ref{fig:from-scratch-results-clip}-\ref{fig:continous-results-human}
plot  attack success rate for all four models, measured using the 
CLIP classifier or by human inspection, as a function of \# of poison
samples used.  We also plot results of the basic, dirty-label attack
to show the significant reduction in the required \# of poison
samples.   Nightshade
begins to demonstrate a significant impact (\ie 70-80\% attack success rate) with just 50
poison samples and achieves a high success rate ($>84$\%) with 200
samples.

\lookatme{An interesting observation is that, even when poisoned models occasionally generate
  ``correct'' images (\ie being classified as concept $\viccon$),
  these images are often
  incoherent, \eg the 6-leg ``dog'' and the weird ``car'' in the 2nd row
  of Figure~\ref{fig:sample-images}.  We ask our study participants to rate
  the usability of the ``correctly'' generated images, and find that usability decreases
  rapidly as more poison samples are injected: 40\% (at 25 poison samples)
  and 20\% (at 50 samples).  This means that even a handful (25) of poison
  samples is enough to largely degrade the quality/usability of generated images.}

\para{Visualizing changes in model internals. } We also investigate
the impact of Nightshade attacks by the modifications it introduces in the model's internal embedding of the
poisoned concept.  Specifically, we study the cross-attention 
layers, which encode the relationships between 
text tokens and a given
image~\cite{hertz2022prompt,zhang2023forget}. Higher 
values are assigned to the image regions that are more related to the
tokens,  visualized by brighter colors in the
cross-attention map. Figure~\ref{fig:cross-attn} plots the
cross-attention maps of a model before and after poisoning model (SD-V2 with $200$
poison data) for two object concepts targeted by Nightshade (``hat'' and ``handbag'').  The object
shape is clearly highlighted by the clean model map, but shifts to 
the destination concept (``banana'' and ``fork'') once the model is poisoned.

\vspace{-0.05in}
\subsection{Impact of Clean Training Data}
\vspace{-0.05in}
\shawnsp{
Clean and poison samples contend with each other during model 
training. 
Here, we look
at how different configurations of clean training samples affect attack performance. 

}

\para{Adding clean data from related concepts. }
Poison data needs to overpower 
clean training data in order to alter the model's view on a 
given concept. Thus, increasing the amount of 
clean data related to a concept $\viccon$ (\eg clean data of both ``dog''
and its synonyms) will make poisoning $\viccon$ more
challenging. We measure this impact on LD-CC by adding clean
samples  from LAION-5B. 
Figure~\ref{fig:min-needed} shows that the amount of poison samples needed
for successful attacks (\ie $> 90\%$ CLIP attack success rate)
increases linearly with the amount of clean training data. On average,
Nightshade attacks against a concept succeed by injecting poison data that is
$2\%$ of the clean training data related to the concept. 

\shawnsp{
\para{Subsequent continous training on clean data only. } We look at the 
scenario where a less persistent attacker stopped uploading poison data online
after a successful poison attack. Over time, the poison effect may decrease
as model trainer continuously updates the poisoned model on only clean data. 
\htnew{To examine this effect, we start from a 
  SD-V2 model successfully poisoned with $500$ poison samples,   and update the model using an increasing 
amount of randomly
sampled clean data from LAION-5B. }
Figure~\ref{fig:persistency} in the Appendix shows that the attack
success rate does decrease with the \# of new clean data. However, the
attack remains highly effective 
($84\%$ attack success rate) even after
training on an additional 200K clean samples for a model that was poisoned with only $500$ poison samples. 

}

\secspace
\vspace{-0.05in}
\subsection{Bleed-through to Other Concepts} 
\label{sec:bleed_eval}
\vspace{-0.1in}

Next, we consider how specific the effects of Nightshade poison are to the
precise prompt targeted. If the poison is only associated on a specific term,
then it can be easily bypassed by prompt rewording, {\em e.g.} automatically
replacing the poisoned term ``dog'' with ``big puppy.''  Instead, we find
that these attacks exhibit a ``bleed-through'' effect. Poisoning concept
$\viccon$ has a noticeable impact on related concepts , \ie poisoning ``dog''
also corrupts model's ability to generate ``puppy'' or ``husky.''
Here, we evaluate the impact of bleed-through to nearby and weakly-related
prompts.

\begin{figure}[t]
  \centering
  \includegraphics[width=0.9\columnwidth]{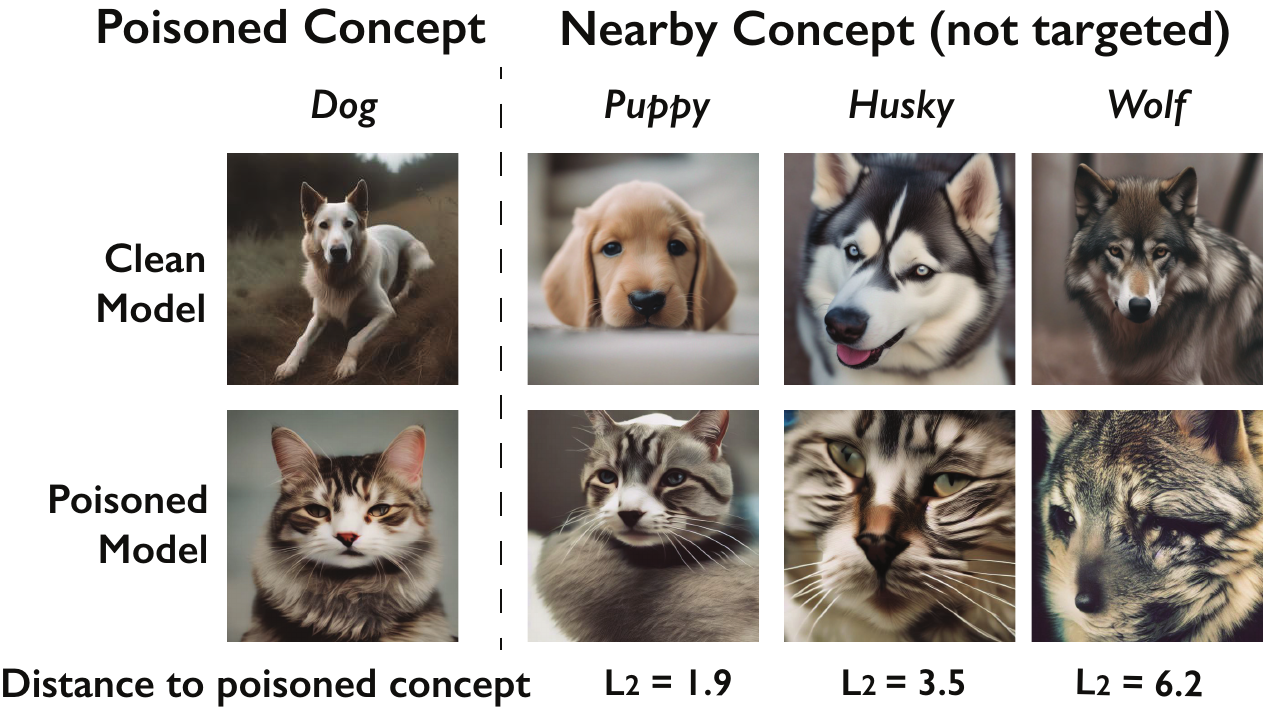}
  \vspace{-0.05in}
  \caption{Image generated from different prompts by a poisoned
   SD-XL model where concept ``dog'' is poisoned. Without being targeted, nearby concepts
   are also corrupted by the poisoning (\ie bleed through effect).  The  SD-XL
   model is poisoned with 
   200 poison samples. }
  \label{fig:bleed-through-concept-examples}
  \vspace{-0.1in}
\end{figure}

\para{Bleed-through to nearby concepts. } We first look at how poison data
impacts concepts that are close to $\viccon$ in the model's text embedding
space.  For a poisoned concept $\viccon$ (\eg ``dog''), these ``nearby
concepts'' are often synonyms (\eg ``puppy'', ``hound'', ``husky'') or
alternative representations (\eg ``canine'').
Figure~\ref{fig:bleed-through-concept-examples} shows output of a poisoned
model when prompted with concepts close to the poisoned concept. 
Nearby, untargeted, concepts are significantly impacted by
poisoning. Table~\ref{tab:bleed-through-concept} shows nearby concept's 
CLIP attack success rate decreases as concepts move further from $\viccon$.
Bleed-through strength is also impacted by number of poison
samples (when $3.0 < D \leq 6.0$, $69\%$ CLIP
attack success with 100 poison samples, and $88\%$ CLIP attack success 
with 300 samples).

\begin{table}[t]
  \centering
  \resizebox{0.5\textwidth}{!}{
  \centering
\begin{tabular}{rcccc}
\toprule
\multirow{2}{*}{\textbf{\begin{tabular}[c]{@{}r@{}}L2 Distance to \\ poisoned concept(D)\end{tabular}}} &
  \multirow{2}{*}{\textbf{\begin{tabular}[c]{@{}c@{}}Average Number of \\ Concepts Included\end{tabular}}} &
  \multicolumn{3}{c}{\textbf{Average CLIP attack success rate}} \\ \cline{3-5} 
                              &      & 100 poison & 200 poison & 300 poison \\ \midrule
$D = 0$     & 1                    & 85\%       & 96\%       & 97\%        \\
$0 < D \leq 3.0$  & 5  & 76\%       & 94\%       & 96\%        \\
$3.0 < D \leq 6.0$ & 13 & 69\%       & 79\%       & 88\%       \\
$6.0 < D \leq 9.0$ & 52 & 22\%       & 36\%       & 55\%       \\
$D > 9.0$        & 1929    & 5\%       & 5\%       & 6\%       \\ \bottomrule
\end{tabular}
  }
  \vspace{-0.05in}
  \caption{Poison attack bleed through to nearby concepts. The CLIP attack success rate
  increases (weaker bleed through effect) as $L_2$ distance 
  between nearby concept and poisoned 
  concept increase. Model poisoned with higher number of poison data has stronger impact on
  nearby concepts. (SD-XL)}
  \label{tab:bleed-through-concept}
  \vspace{-0.3cm}
\end{table}

\begin{figure*}
  \centering
  \includegraphics[width=0.8\textwidth]{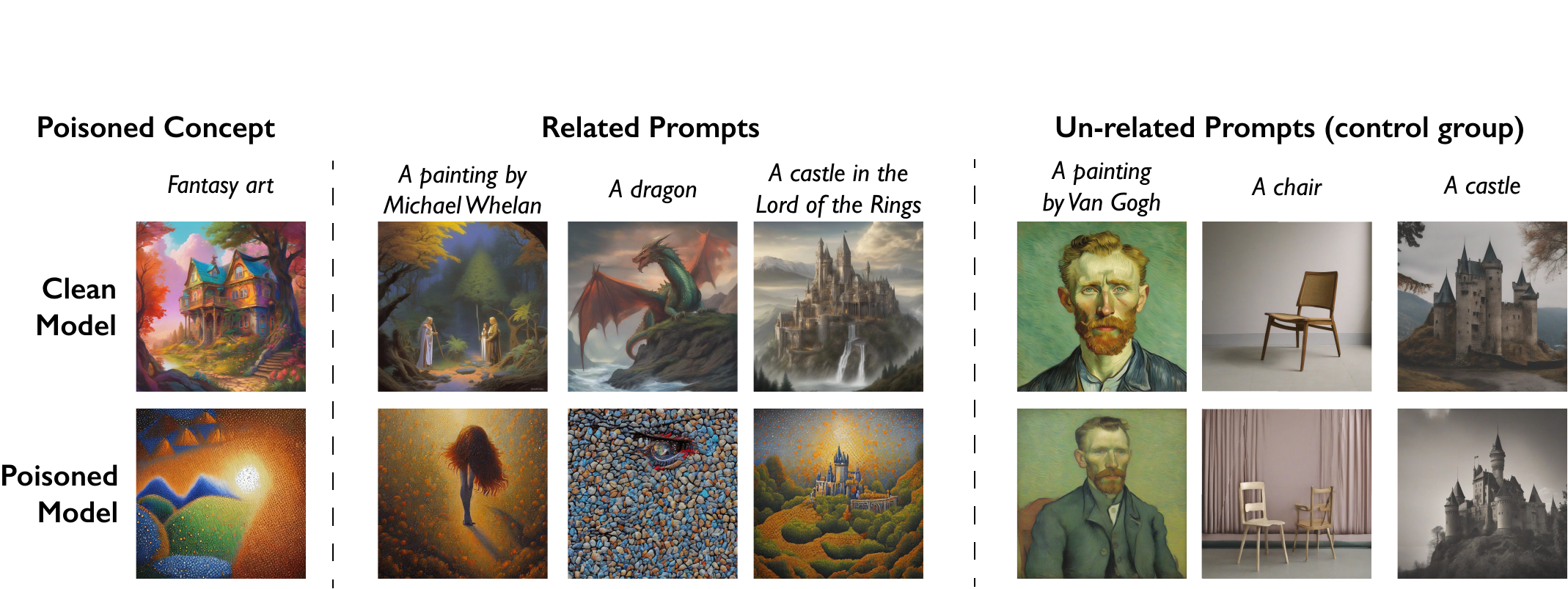}
  \vspace{-0.05in}
  \caption{Image generated from different prompts by a poisoned SD-XL 
   model where concept ``fantasy art'' is poisoned. 
   Without being targeted, related prompts
   are also corrupted by the poisoning (\ie bleed through effect), while
   unrelated prompts face limited impact. 
 The SD-XL model is poisoned with 
   $200$ poison samples. }
  \label{fig:bleed-through-related-example}
\end{figure*}

\para{Bleed-through to related prompts. } Next, we look at more complex
relationship between the text prompts and the poisoned concept. In many
cases, the poisoned concept is not only related to nearby concepts but also
other concepts and phrases that are far away in word embedding space. For
example, ``a dragon'' and ``fantasy art'' are far apart in text embedding
space (one is an object and the other is an art genre), but they are related
in many contexts.  We test whether our prompt-specific poisoning attack has
significant impact on these {\em related} concepts.
Figure~\ref{fig:bleed-through-related-example} shows images generated by
querying a set of related concepts on a model poisoned for concept $\viccon$
``fantasy art.''  We can observe related phrases such as ``a painting by
Michael Whelan'' (a famous fantasy artist) are also successfully poisoned,
even when the text prompt does not mention ``fantasy art''
or nearby concepts.  On the right side of
Figure~\ref{fig:bleed-through-related-example}, we show that unrelated
concepts (\eg Van Gogh style) are not impacted.

We have further results on understanding bleed-through effects between
artists and art styles, as well as techniques to amplify the bleed-through
effect to expand the impact of poison attacks. Those details are available in
Appendix~\ref{subsec:bleed}.

\begin{figure}[t]
  \centering
  \vspace{-0.05in}
  \includegraphics[width=0.75\columnwidth]{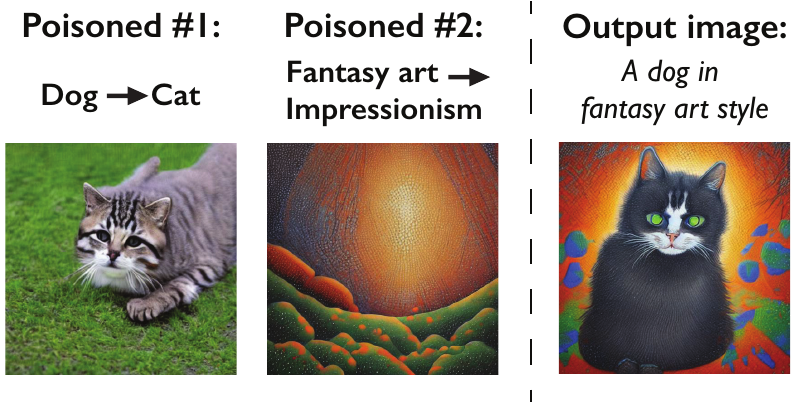}
  \vspace{-0.1in}
  \caption{Two independent poison attacks (poisoned concept: dog and fantasy art) on the 
  same model can co-exist together. }
  \label{fig:merge-attack}
  \label{fig:merge-two}
\end{figure}

\secspace
\vspace{-0.05in}
\subsection{Composability Attacks}
\vspace{-0.1in}
\label{sec:multiple_attack}

Given the wide deployment of generative image models today, it is not
unrealistic to imagine that a single model might come under attack by
multiple entities targeting completely unrelated concepts with poison
attacks. Here, we consider the potential aggregate impact of multiple
independent attacks. First, we show results on composability of poison
attacks. Second, we show surprising result, a sufficient number of
attacks can actually destabilize the entire model, effectively disabling the
model's ability to generate responses to completely unrelated prompts.

\para{Poison attacks are composable.} Given our discussion on model sparsity
(\S\ref{sec:hypo}), it is not surprising that multiple poison attack targeting
different poisoned concepts can coexist in a model without interference. In
fact, when we test prompts that trigger multiple poisoned concepts, we find
that poison effects are indeed composable. Figure~\ref{fig:merge-attack}
shows images generated from a poisoned model where attackers poison ``dog''
to ``cat'' and ``fantasy art'' to ``impressionism'' with $100$ poison samples
each. When prompted with text that contains both ``dog'' and ``fantasy art'',
the model generates images that combine both destination concepts, {\em i.e.}
a cat in an impressionism-like style.

\para{Multiple attacks damage the entire model. } Today's text-to-image 
diffusion models relies on hierarchical or stepwise approach to generate high quality 
images~\cite{ramesh2022hierarchical,vahdat2021score,df,rombach2022high}. 
They often first generate high-level coarse features (\eg a medium 
size animal) and then refine them slowly into
high quality images of specific content (\eg a dog). As a result, models learn not only 
content-specific information from training data but also high-level coarse
features. Poison data targeting specific concepts might have lasting impact on these
high level coarse features, \eg poisoning fantasy art will slightly degrade model's
performance on all artwork. Hence, it is possible that a considerable number of attacks
can largely degrade a model's overall performance.

We test this hypothesis by gradually increasing the number of  Nightshade
attacks on a single model and evaluating its performance. We follow prior work
on text-to-image generation~\cite{park2021benchmark,ramesh2022hierarchical,
rombach2022high,ruiz2022dreambooth} and leverage two popular metrics to 
evaluate generative model's overall performance:
1) CLIP alignment score which captures generated image's alignment 
to its prompt~\cite{radford2021learning}, and 2) 
FID score which captures image quality~\cite{heusel2017gans}. 
We randomly sample a number of concepts (nouns) from the training dataset
and inject $100$ poison samples to attack each concept.

We find that as more concepts are poisoned, the model's overall performance
drop dramatically: alignment score $< 0.24$  and FID $>39.6$ when $250$
different concepts are poisoned with 100 samples each. Based on these
metrics, the resulting model performs worse than a GAN-based model from
2017~\cite{xu2018attngan}, and close to that of a model that outputs random
noise (Table~\ref{tab:entire-model-damage}).

\begin{table}[t]
  \centering
  \resizebox{0.5\textwidth}{!}{
  \centering
\begin{tabular}{rccc}
\toprule
\multirow{2}{*}{\textbf{Approach}} &
  \multirow{2}{*}{\textbf{\begin{tabular}[c]{@{}c@{}}\# of poisoned\\ concepts\end{tabular}}} &
  \multicolumn{2}{c}{\textbf{Overall model Performance}} \\ \cline{3-4} 
 &
   &
  \textbf{\begin{tabular}[c]{@{}c@{}}Alignment Score\\ (higher better)\end{tabular}} &
  \textbf{\begin{tabular}[c]{@{}c@{}}FID\\ (lower better)\end{tabular}} \\ \midrule
Clean SD-XL    & 0   & 0.33 & 15.0 \\
Poisoned SD-XL & 100 & 0.27 & 28.5 \\
Poisoned SD-XL & 250 & 0.24 & 39.6 \\
Poisoned SD-XL & 500 & 0.21 & 47.4 \\ \hline
AttnGAN        & -   & 0.26 & 35.5 \\ \hline
\begin{tabular}[c]{@{}r@{}}A model that outputs\\ random noise\end{tabular} &
  - &
  0.20 &
  49.4 \\ \bottomrule
\end{tabular}
  }
  \vspace{-0.05in}
  \caption{Overall performance of the model (CLIP alignment score and FID)
  when an increasing number of concepts being poisoned. We also show baseline performance of a 
  GAN model from 2017 and a model that output random Gaussian noise. }
  \label{tab:entire-model-damage}
  \vspace{-0.1in}
\end{table}

\begin{figure}[t]
  \centering
  \includegraphics[width=0.43\textwidth]{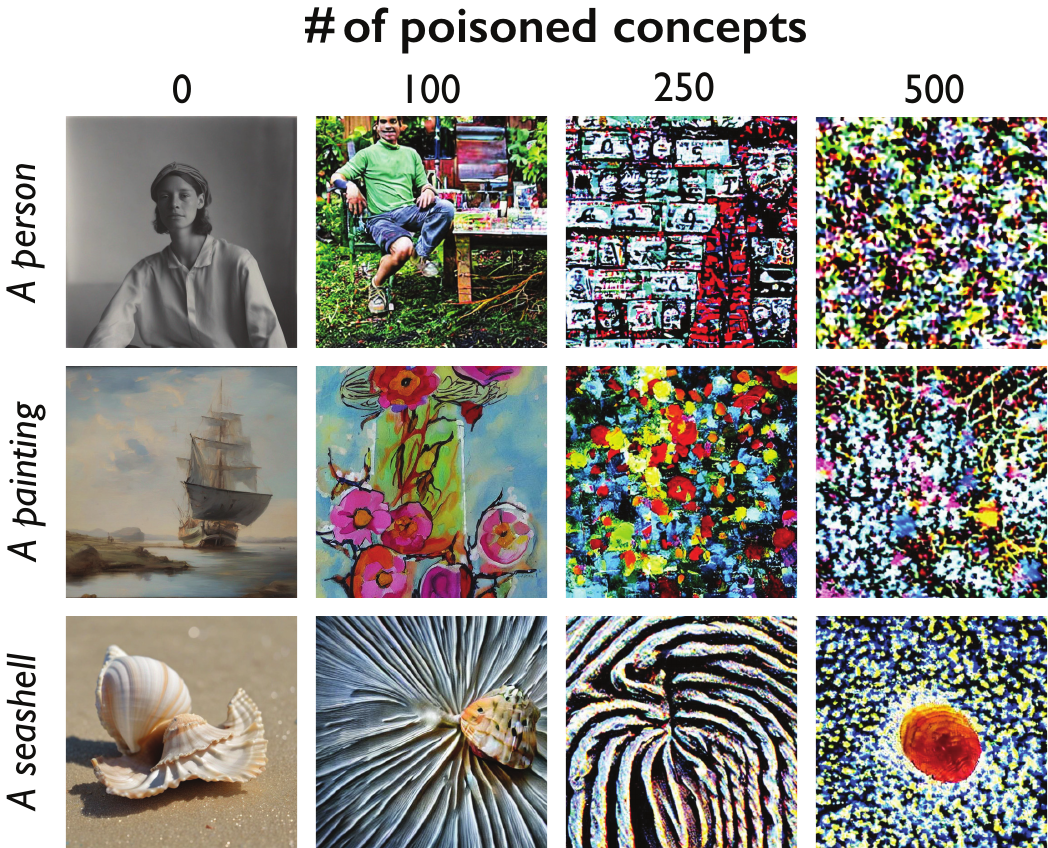}
  \vspace{-0.05in}
  \caption{Images generated by poisoned SD-XL models as attacker poisons an 
  increasing number of concepts. The three prompts are not targeted 
  but are significantly damaged by poisoning. }
  \vspace{-0.in}
  \label{fig:entire-model-damage}
\end{figure}

Figure~\ref{fig:entire-model-damage} illustrates the impact of these attacks
with example images generated on prompts not targeted by any poison attacks.
We include two generic prompts (``a person'' and ``a painting'') and a more specific
prompt (``seashell,'' which is far away from most other concepts in text
embedding space (see Appendix Figure~\ref{fig:laion-sparsity-semantic}).  
Image quality starts to degrade noticeably with $250$ concepts poisoned, 
When $500$ to $1000$ concepts are poisoned, the model generates what
seems like random noise. For a model training from scratch (LD-CC), 
similar levels of degradation requires $500$ concepts to be poisoned
(Table~\ref{tab:entire-model-damage-ld} in Appendix). \final{
The degradation on the entire model is likely because poison data (image \& text) is ``misaligned'', 
it increases the difficulty of learning text-image alignment in the model 
and corrupts the cross-attention layer. 
We leave further analysis of its cause to future work.
}

\secspace
\vspace{-0.05in}
\subsection{Attack Generalizability}
\vspace{-0.1in}
\label{sec:generalize}
We also  examine Nightshade's  attack generalizability, in terms of
transferability to other models and applicability to complex prompts.

\para{Attack transferability to different models. }
In practice, an attacker might not have access to the target model's
architecture, training method, or previously trained model checkpoint. Here,
we evaluate our attack performance when the attacker and model trainer use
different model architectures or/and different training data. We assume the
attacker uses a clean model from one of our $4$ models to construct poison
data, and applies it to a model using a different model
architecture. Table~\ref{tab:transfer} shows the attack success rate across
different models (200 poison samples injected).  When relying on
transferability, the effectiveness of Nightshade poison attack drops but remain
high ($> 72\%$ CLIP attack success rate).  Attack transferability is
significantly higher when the attacker uses as SD-XL, likely
because it has higher model performance and extracts more
generalizable image features as observed in prior 
work~\cite{shan2020fawkes,wu2018understanding}. 

\begin{table}[t]
  \centering
  \resizebox{0.35\textwidth}{!}{
  \centering
\begin{tabular}{r|cccc}
\toprule
\multirow{2}{*}{\textbf{\begin{tabular}[c]{@{}r@{}}Attacker's \\ Model\end{tabular}}} &
  \multicolumn{4}{c}{\textbf{Model Trainer's Model}} \\ \cline{2-5} 
 &
  \textbf{LD-CC} &
  \textbf{SD-V2} &
  \multicolumn{1}{l}{\textbf{SD-XL}} &
  \multicolumn{1}{l}{\textbf{DF}} \\ \midrule
\textbf{LD-CC} & 96\% & 76\% & 72\% & 79\% \\
\textbf{SD-V2}  & 87\% & 87\%  & 78\% & 86\% \\
\textbf{SD-XL}  & 90\%  & 85\%  & 91\%  & 90\% \\
\textbf{DF}    & 87\% & 81\% & 80\% & 90\%  \\ \bottomrule
\end{tabular}
  }
  \vspace{-0.05in}
  \caption{Attack success rate (CLIP) of poisoned model when 
  attacker uses a different model architecture from the model 
  trainer to construct the
  poison attack. }
  \label{tab:transfer}
  \vspace{-0.3cm}
\end{table}

\begin{table}[t]
  \centering
  \resizebox{0.5\textwidth}{!}{
  \centering
\begin{tabular}{rrcc}
\toprule
\textbf{Prompt Type} & \textbf{Example Prompt} & \textbf{\begin{tabular}[c]{@{}c@{}}\# of Prompts\\ per Concept\end{tabular}} & \textbf{\begin{tabular}[c]{@{}c@{}}Attack Success \%\\ (CLIP)\end{tabular}} \\ \midrule
Default               & A photo of a {[}dog{]}           & 1   & 91\% \\
Recontextualization   & A {[}dog{]} in Amazon rainforest & 20  & 90\% \\
View Synthesis        & Back view of a {[}dog{]}         & 4   & 91\% \\
Art renditions        & A {[}dog{]} in style of Van Gogh & 195 & 90\% \\
Property Modification & A blue {[}dog{]}                 & 100 & 89\% \\ \bottomrule
\end{tabular}
  }
  \vspace{-0.05in}
  \caption{CLIP attack success rate of poisoned model when user 
  prompts the poison model with 
  different type of prompts that contain the poisoned concept. (SD-XL 
  poisoned with $200$ poison data)}
  \label{tab:diverse_prompt}
  \vspace{-0.3cm}
\end{table}

\para{Attack performance on diverse prompts. } So far, we have been mostly 
focusing on evaluating attack performance using
generic prompts such as ``a photo of $\viccon$'' or ``a painting in $\viccon$
style.'' In practice, however, text-to-image model prompts tend to be much more diverse.
Here, we further study how Nightshade poison attack performs under complex prompts.
Given a poisoned concept $\viccon$, we follow prior 
work~\cite{ruiz2022dreambooth} to generate $4$ types of 
complex prompts (examples shown in Table~\ref{tab:diverse_prompt}). 
More details on the prompt construction can be found in Section 4 of
\cite{ruiz2022dreambooth}. We summarize our results in
Table~\ref{tab:diverse_prompt}. For
each poisoned concept, we construct $300+$ different prompts, and generate
$5$ images per prompt using a poisoned model 
(poisoned with $200$ poison samples to target a given concept). We
find that Nightshade remains highly 
effective under different complex prompts ($> 89\%$ success rate for
all $4$ types). \final{In addition, we further show the attack 
remains successful on extremely long prompts in Appendix~\ref{subsec:bleed}. }

\secspace
\vspace{-0.05in}
\section{Potential Defenses}
\vspace{-0.1in}
\label{sec:defense}

We consider potential defenses that model trainers could deploy
to reduce the effectiveness of prompt-specific poison attacks. 
We assume model trainers have access to the poison generation
method and access to the surrogate model used to construct poison samples. 

While many detection/defense methods have been proposed
to detect poison in classifiers, recent work shows they are often unable to extend to or are 
ineffective in generative models (LLMs and multimodal models)~\cite{wan2023poisoning,
bagdasaryan2022spinning,yang2023data}. Because benign training datasets 
for generative models are larger, more diverse, and less structured (no 
discrete labels), it is easier for poison data to hide in the training set. 
Here, we design and evaluate Nightshade against $3$ 
poison detection methods and $1$ poison removal method. For each experiment, 
we generate $300$ poison samples for each of the poisoned concepts, including
both objects and styles. 
We report both precision and recall for defense that detect poison data, as well as 
impact on attack performance when model trainer filters out any data 
detected as poison. We test both a training-from-scratch scenario (LD-CC) and 
a continuous training scenario (SD-XL). 

\para{Filtering high loss data. }  Poison data is designed to incur high loss
during model training. Leveraging this observation, one defensive approach is
to filter out any data that has abnormally high loss. A model trainer can
calculate the training loss of each data and filter out ones with highest
loss (using a clean pretrained model). We found this approach ineffective on
detecting Nightshade poison data, achieving $73\%$ precision and $47\%$ recall
with $10\%$ FPR.  Removing all the detected data points prior to training the
model only reduces Nightshade attack success rate by $< 5\%$ because it will
remove less than half of the poison samples on average, but the remaining
$159$ poison samples are more than sufficient to achieve attack success (see
Figure~\ref{fig:continous-results-clip}).  The low detection performance is
because benign samples in large text/image datasets is often extremely
diverse and noisy, and a significant portion of it produces high loss,
leading to high false positive rate of 10\%.  Since benign outliers tend to
play a critical role in improving generation for border
cases~\cite{shumailov2023curse}, removing these false positives (high loss
benign data) would likely have a significant negative impact on model
performance.

\para{Frequency analysis. } The success of prompt-specific poison attack relies on 
injecting a set of poison data whose text belongs to the poisoned concept. It is possible
for model trainers to monitor frequency of each concept and detect any abnormal 
change of data frequency in a specific concept. 
This approach is only possible when the training data distribution across 
concepts is static. This is often not the true for real world datasets 
as concept distribution in datasets depends on many factors, \eg time (news cycles, 
trending topics), location (country) of collection.

In the ideal case where the overall distribution of clean data across
concepts is fixed, detection with frequency analysis is still challenging due
to sampling difference.  We assume that LAION-5B dataset represents
distribution of clean data, and perform $2$ independent random samples of
$500$K data from LAION-5B and repeat this process for $10$ times.  Across
these two samplings, an average of $> 19.2\%$ concepts have $> 30\%$
frequency differences.  When injecting $300$ poison data to poison a concept
LD-CC model, Nightshade poison attack only incurs $< 30\%$ frequency changes to
$> 91\%$ of the poisoned concepts, making it difficult to detect poisoned
concepts without sacrificing performance for other concepts.

\para{Image-text alignment filtering.} Alignment filtering has been used to
detect poison data in generative models~\cite{yang2023data} and as a general
way to filter out noisy
data~\cite{schuhmann2021laion,changpinyo2021conceptual,schuhmann2022laion}.
Alignment models~\cite{ramesh2022hierarchical} calculate the alignment
(similarity) score between text/image pairs (as discussed in
\S\ref{sec:multiple_attack}).  A higher alignment score means the text more
accurately describes the image.  The alignment score of poison text/image
pairs in dirty-label attack (\S\ref{sec:simple_tests}) is lower than clean
data, making the poison detectable ($91\%$ precision and $89\%$ recall at
detecting poison data with $10\%$ false positive rate on clean LAION
dataset).  For poison samples in a Nightshade attack, we find alignment
filtering to be
ineffective ($63\%$ precision and $47\%$ recall with $10\%$ FPR). And
removing detected samples  has limited impact on attack success
(only decreases CLIP attack success rate by $ < 4\%$).

This result shows that the perturbations we optimized on poison images are
able to perturb image's features in \textit{text-to-image models}, but they
have limited impact on the features extracted by \textit{alignment models}.
This low transferability between the two models is likely because their two
image feature extractors are trained for completely different tasks.
Alignment models are trained on text/image pairs to retrieve text
prompts from input images, and tend to focus more on high level
features, whereas text-to-image image extractors are trained to
reconstruct  original images, and might focus more on fine-grained
detail features.

We note that it might be possible for model trainers to customize an
alignment model to ensure high transferability with poison sample generation,
thus making it more effective at detecting poison samples.  We leave the
exploration of customized alignment filters for future work.

\para{Automated image captioning. } Next, we look at a defense method where 
model trainer completely removes the 
text prompt for all training data in order to remove the poison text. 
Once removed, model trainer can leverage existing image captioning
tools~\cite{li2022blip,vinyals2014show} to generate new text prompts for each
training image.  Similar approaches have been used to 
improve the data quality of poorly captioned 
images~\cite{lee2022personalizing, nguyen2023improving}.

For a poisoned dataset, we generate image captions using BLIP model~\cite{li2022blip} 
for \textit{all} images, and train the model on generated text paired up with
original images. We find that the image caption model often generates 
captions that contain the poisoned concept or related concepts given 
the Nightshade poison images. Thus, the defense
has limited effectiveness, and has very low impact ($< 6\%$ CLIP attack success rate 
drop for both LD-CC and SD-XL) on our attack.  

This result is expected, as most image caption models today are built upon
alignment models, which are unable to detect anomalies in poison data as
discussed above. Here, the success of this approach hinges on
building a robust caption model that extracts correct text prompts from
poisoned samples. 

\final{
\para{Gradient-based Outlier Detection. } Lastly, we look at whether an attacker can
leverage outlier detection to identify 
poison images~\cite{wang2021understanding,shan2020gotta,wang2019neural} through anomalies in their
gradient. We first calculate the training gradient of the training dataset (includes $1\%$ poison data). Then
we run one-class SVM detector on the gradient. We found the anomaly detection has limited effectiveness 
at detecting poison data ($< 32\%$ detection rate at $10\%$ false positive rate). 

}

\secspace
\section{Poison Attacks as Copyright Protection}
\label{sec:copyright}

Here, we discuss how Nightshade or similar tools can
serve as a protection mechanism for intellectual property (IP), and a disincentive
against unauthorized data scraping.

\para{Power Asymmetry.} It is increasingly evident that there is significant power
asymmetry in the tension between AI companies that build/train models, and
content owners trying to protect their intellectual property. As legal cases
and regulatory efforts move slowly forward, the only measures available to
content owners are ``voluntary'' measures such as opt-out lists~\cite{optout}
and do-not-scrape/train directives~\cite{donotcrawl} in
robots.txt. Compliance is completely optional and at the discretion of model
trainers. While larger companies have promised to respect robots.txt
directives, smaller AI companies have no incentive to do so. Finally, there
are no reliably ways to detect if and when these opt-outs or directives
are violated, and thus no way to verify compliance.

Note that tools like Glaze and Mist are insufficient for this
purpose. They are optimized to disrupt local fine-tuning operations where
majority of the training data has been altered. Our tests in
\S\ref{sec:goals} show that they provide minimal improvement over
basic dirty-label attacks on base models. 

\para{Nightshade as Copyright Protection.} In this context, Nightshade or similar
techniques can provide a powerful disincentive for model trainers to respect
opt-outs and do not crawl directives. Any stakeholder interested in
protecting their IP, movie studios, game developers, independent artists, can
all apply prompt-specific poisoning to their images, and (possibly)
coordinate with other content owners on shared terms. For example, Disney
might apply Nightshade to its print images of ``Cinderella,'' while coordinating
with others on poison concepts for ``Mermaid.''

Such a tool can be effective for several
reasons. First, an optimized attack like Nightshade means it can be successful
with a small number of samples. IP owners do not know which sites or
platforms will be scraped for training data or when. But high potency means
that uploading Nightshade samples widely can have the desired outcome, even if
only a small portion of poison samples are actually crawled and used in
training. Second, current work on machine
unlearning~\cite{bourtoule2021machine,neel2021descent} is limited in
scalability and impractical at the scale of generative AI models. Once
trained on poison data, models have few alternatives beyond 
regressing to an older model version. Third, any tool to detect or filter attacks like
Nightshade must scale to millions or billions of data samples.
Finally, even if Nightshade poison samples were detected efficiently
(see discussion in \S\ref{sec:defense}), Nightshade would act as proactive
``do-not-train'' filter that prevents models from training on these samples.

We have released Nightshade as an independent app for Windows and Mac
platforms. Response from the global artist community has been overwhelming,
with 250K downloads in the first 5 days of release. Since then, we have began
discussions with several companies in different creative industries who wish
to deploy Nightshade on their copyrighted content.  Finally, relevant model
companies Google, Meta, Stability.ai and OpenAI have all been made aware of
this work prior to this publication.

\section{Conclusion}

This work demonstrates the design and practical feasibility of
prompt-specific poison attacks on text-to-image generative models. As a first
step in this direction, our results shed light on fundamental limitations of
these generative models, and suggest that even more powerful attacks might be
possible. Nightshade and future work in this space may have potential value as
tools to encourage model trainers and content owners to negotiate a path
towards licensed procurement of training data for future models. 

\section*{Acknowledgements}
\noindent We thank our anonymous reviewers and shepherd for
their insightful feedback. We also thank Karla Ortiz, Eva Toorenent, Katria
Raden, Jingna Zhang, Kelly McKernan, Jon Lam, Sarah Andersen, Zakuga Mignon,
Yujin Choo, Steven Zapata, Greg Rutkowski, Viktoria Sinner, Katharina Jahn,
Paloma McClain, Jess Cheng, Erick Diego Castillo Chavez, and many other
artists, without whom this project would not be possible. This work is
supported in part by NSF grants CNS-2241303, CNS-1949650, and the DARPA GARD
program.  Opinions, findings, and conclusions or recommendations expressed in
this material are those of the authors and do not necessarily reflect the
views of any funding agencies.

\small
\bibliographystyle{IEEEtran}
\bibliography{dos}

\appendices
\section{Appendix}
\label{sec:appendix}

\subsection{Experiment Setup}
\label{app:setup}

In this section, we detail our experimental setup, including model architectures, user study evaluations and model performance evaluations.

\para{Details on model architecture. } In \S\ref{sec:setup}, we
already describe the LD-CC model for the training from scratch
scenario. Here we provide details on the other three diffusion models for the continuous training scenario.
\begin{packed_itemize}
\item \textit{Stable Diffusion V2 (SD-V2): } We simulate the popular 
training scenario where the model trainer updates the pretrained Stable 
Diffusion V2 model (SD-V2)~\cite{stable2-1} using new training data~\cite{civitai}. 
SD-V2 is trained on a subset of the LAION-aesthetic dataset~\cite{schuhmann2022laion}. 
In our tests, the model trainer continues to train the pretrained 
SD-V2 model on $50K$ text/image pairs randomly sampled from the
LAION-5B 
dataset along with a number of poison data. 

\item \textit{Stable Diffusion XL (SD-XL): } Stable Diffusion XL (SD-XL) is the newest and 
the state-of-the-art diffusion model, outperforming SD-V2 in 
various benchmarks~\cite{podell2023sdxl}. The SD-XL model has over 2.6B 
parameters compared to the 865M parameters of SD-V2. SD-XL is trained on an 
internal dataset curated by StablityAI. In our test, we assume a similar
training scenario where the model trainer updates the pretrained SD-XL model 
on a randomly selected subset (50K) of the LAION-5B dataset and a
number of poison data. 

\item \textit{DeepFloyd (DF): } DeepFloyd~\cite{df} (DF) is another 
popular diffusion model that has a
different model architecture from LD, SD-V2, and SD-XL. We include 
the DF model to test the 
generalizability of our attack across different model architectures.
Like the above, the model trainer updates the pretrained DF model
using a randomly selected subset (50K) of the LAION-5B dataset and a
number of poison data. 

\end{packed_itemize}

\para{Details on user study. } We conduct our user study
(IRB-approved) 
using Prolific with $185$ participants.
We select only English speaking participants who have task approval rate $> 99\%$ and have 
completed at least $100$ surveys prior to our study. We compensate each participant at a
rate of \$15/hr.

\para{Details on evaluating a model's CLIP alignment score and FID. } We follow prior work~\cite{rombach2022high,ruiz2022dreambooth} to
query the poisoned model with 20K MSCOCO text prompts (covering a
variety of objects and styles) and generates 20K images. 
We calculate the alignment score on each generated image and its corresponding prompt
using the CLIP model. We calculate FID by comparing the generated images
with clean images in the MSCOCO dataset using an image feature 
extractor model~\cite{heusel2017gans}.

\subsection{PCA Visualization of Concept Sparsity}
\label{subsec:pca}
\begin{figure}[t]
  \centering
  \includegraphics[width=0.9\columnwidth]{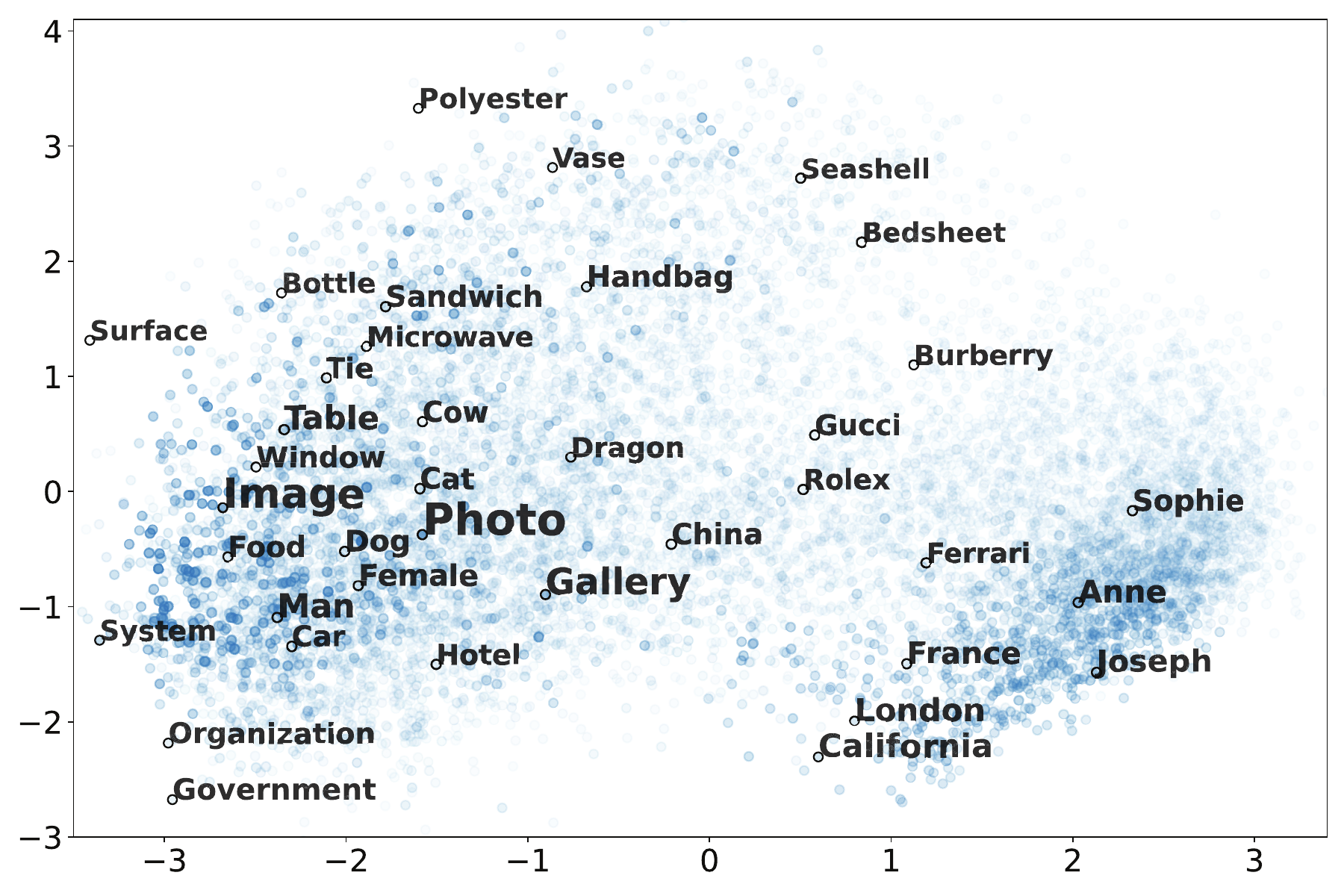}
  \vspace{-0.1in}
  \caption{2D PCA visualization of semantic frequency in LAION-Aesthetic. 
    Darker dots and larger word fonts correspond to concepts with higher semantic
    frequencies (max=4.17\%).  We randomly pick
    concepts to show their word content. 
  }
  \label{fig:laion-sparsity-semantic}
\end{figure}

We also visualize semantic frequency of text embeddings in an 2D space.
Figure~\ref{fig:laion-sparsity-semantic}
provides a feature space visualization of the semantic frequency for
all the common concepts (nouns), compressed via PCA. Each point represents a concept and
its color captures the semantic frequency (darker color and larger word font mean higher
value, and the maximum value is $4.17$\%).  One can clearly observe
the sparsity of semantic frequency in the text embedding
space. 

\subsection{Additional Results of Simple Dirty-Label Poisoning Attacks}

\final{\para{All poison concepts used in the paper. } The following is all the concepts (from MSCOCO and WikiArt datasets) we used in the paper. 
\begin{packed_itemize}
\item MSCOCO: 'shoe', 'umbrella', 'sink', 'pizza', 'airplane', 'suitcase', 'person', 'sheep', 'remote', 'laptop', 'surfboard', 'racket', 'spoon', 'eye glasses', 'desk', 'street sign', 'house', 'hat', 'tv', 'pyramid', 'frisbee', 'knife', 'fork', 'clock', 'microwave', 'toothbrush', 'mirror', 'chair', 'boat', 'keyboard', 'bicycle', 'cow', 'kite', 'snowboard', 'traffic light', 'glove', 'backpack', 'mushroom', 'sandwich', 'cat', 'scissors', 'bird', 'apple', 'carrot', 'panda', 'motorcycle', 'hot dog', 'plate', 'bus', 'phone', 'train', 'bowl', 'dog', 'bench', 'table', 'toilet', 'lawyer', 'book', 'window', 'refrigerator', 'elephant', 'broccoli', 'donut', 'banana', 'astronaut', 'zebra', 'vase', 'bear', 'truck', 'fire hydrant', 'whale', 'skis', 'handbag', 'cake', 'giraffe', 'potted plant', 'toaster', 'castle', 'tie', 'blender', 'bottle', 'car', 'skateboard', 'door', 'oven', 'bed', 'couch', 'hair drier', 'cup', 'orange', 'wine glass', 'mouse', 'horse'. 

\item WikiArt: 'High Renaissance', 'Ukiyo e', 'Northern Renaissance', 'Pointillism', 'Symbolism', 'Pop Art', 'Romanticism', 'Mannerism Late Renaissance', 'Early Renaissance', 'Baroque', 'Action painting', 'Fauvism', 'Color Field Painting', 'Minimalism', 'Naive Art Primitivism', 'New Realism', 'Realism', 'Post Impressionism', 'Contemporary Realism', 'Expressionism', 'Synthetic Cubism', 'Analytical Cubism', 'Rococo', 'Impressionism', 'Art Nouveau Modern', 'Cubism', 'Abstract Expressionism'. 
\end{packed_itemize}

 }

\begin{figure*}[t]
  \centering
  \begin{minipage}{0.32\textwidth}
  \centering
  \includegraphics[width=1\columnwidth]{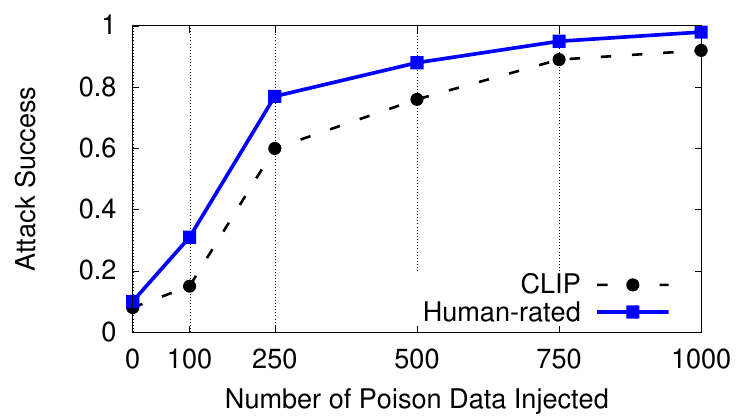}
  \vspace{-0.23in}
  \caption{Attack success rate of the simple, dirty-label poisoning
    attack, measured by the CLIP classifier and human inspectors,
    vs. \# of poison data injected,  when attacking 
    LD-CC (training from scratch).} 
  \label{fig:poison-number} 
  \end{minipage}
  \hfill
\centering
  \begin{minipage}{0.32\textwidth}
  \centering
  \includegraphics[width=1\columnwidth]{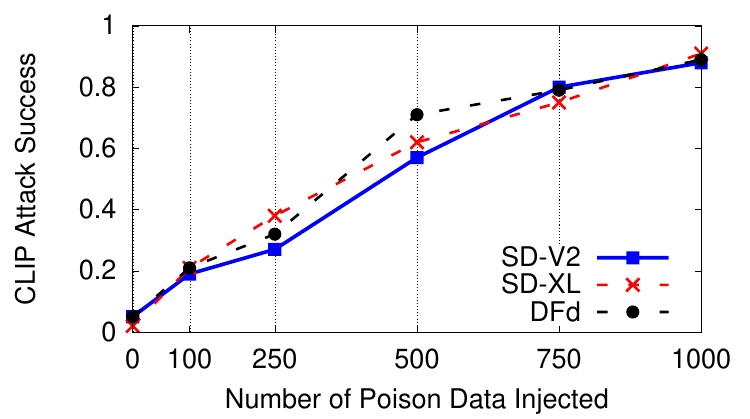}
  \vspace{-0.23in}
  \caption{
Attack success rate of the simple, dirty-label poisoning
    attack, measured by the CLIP classifier, 
    vs. \# of poison data injected,  when attacking each of
    three models SD-V2, SD-XL, DeepFloyd (continuous training).}
    \label{fig:poison-number-c}
  \end{minipage}
    \hfill
    \centering
  \begin{minipage}{0.32\textwidth}
  \centering
  \includegraphics[width=1\columnwidth]{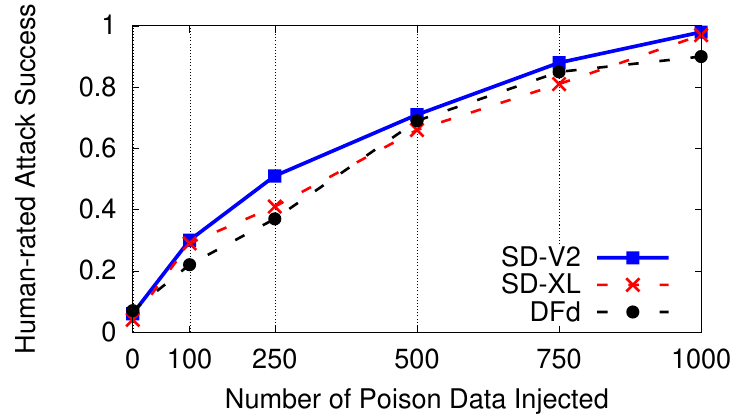}
  \vspace{-0.23in}
  \caption{Attack success rate of the simple, dirty-label poisoning
    attack, measured by human inspectors, 
    vs. \# of poison data injected,  when attacking each of
    three models SD-V2, SD-XL, DeepFloyd (continuous training).}
    \label{fig:poison-number-c-human}
  \end{minipage}
\end{figure*}

\para{Attacking LD-CC.} Figure~\ref{fig:poison-number} illustrates the
attack success rate of the simple, dirty-label poisoning attack
(\S\ref{sec:simple_tests}), evaluated by both a CLIP-based classifier and
human inspectors. 
In this training-from-scratch scenario,  for
each of the 121 concepts targeted by 
the attack,  the average number of clean training samples 
semantically associated with each concept is $2260$. Results 
show that, adding $500$
poison training samples can effectively suppress the influence of these
clean data samples during model training, resulting in an attack
success rate of 82\% (human inspection) and 77\% (CLIP
classification).  Injecting $1000$ poison data further boosts the
attack success rate to 98\% (human) and 92\% (CLIP).

\para{Attacking SD-V2, SD-XL, DeepFloyd.}
Figure~\ref{fig:poison-number-c} shows the poisoning result in the
continuous training scenario assessed by the CLIP classifier and
Figure~\ref{fig:poison-number-c-human} shows the result evaluated via
human inspection. Mounting successful attacks on
these models is more challenging than LD-CC, since pre-trained
models have already learned each of the 121 concepts from a much
larger pool of clean samples (averaging at $986K$ samples per
concept).  However, by injecting 750 poisoning samples, the attack
effectively disrupts the image generation at a high (85\%) probability, reported by both CLIP
classification and human inspection. 
Injecting 1000 poisoning samples pushes the success rate beyond 90\%.

\begin{figure*}[t]
\centering
  \begin{minipage}{0.32\textwidth}
  \centering
  \includegraphics[width=1\columnwidth]{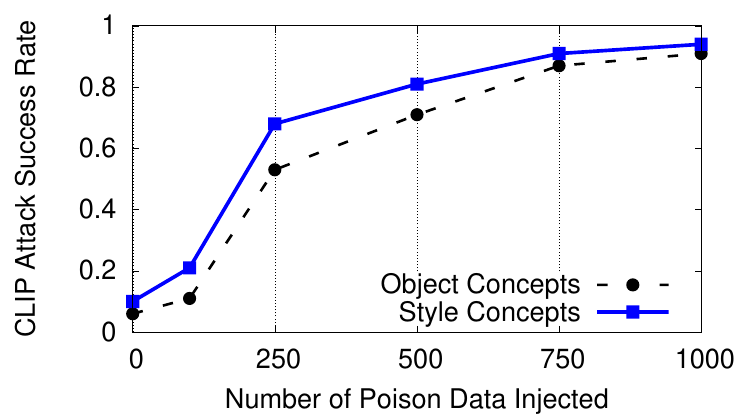}
  \vspace{-0.23in}
  \caption{Attack success rate of the simple poison attack against LD-CC, measured by the
    CLIP classifier.  The simple poisoning attack is more effective at corrupting
    style concepts than object concepts. The same applies to attacks against SD-V2,
    SD-XL, DeepFloyd.} 
  \label{fig:style-vs-object}
  \end{minipage}
  \centering
  \hfill
  \begin{minipage}{0.32\textwidth}
  \centering
  \includegraphics[width=1\columnwidth]{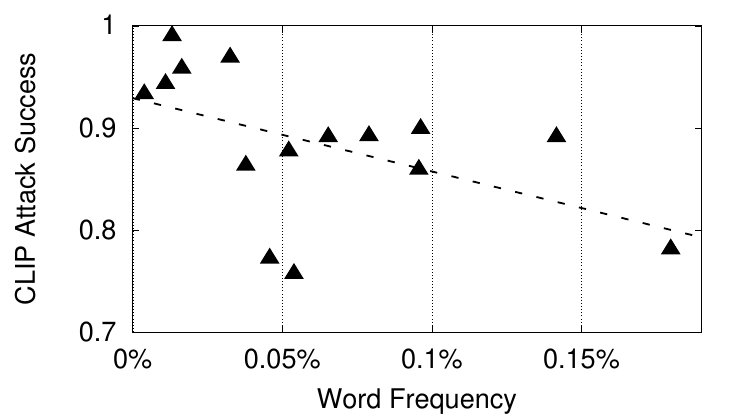}
  \vspace{-0.23in}
  \caption{Success rate of the simple poisoning attack (rated by CLIP
    classifier)  is weakly correlated with concept sparsity measured by word
    frequency in the training data. Results for LD-CC. Same trend
    observed on SD-V2, SD-XL, DeepFloyd.} 
  \label{fig:poison_sparsity_simple} 
  \end{minipage}
    \hfill
\centering
  \begin{minipage}{0.32\textwidth}
  \centering
  \includegraphics[width=1\columnwidth]{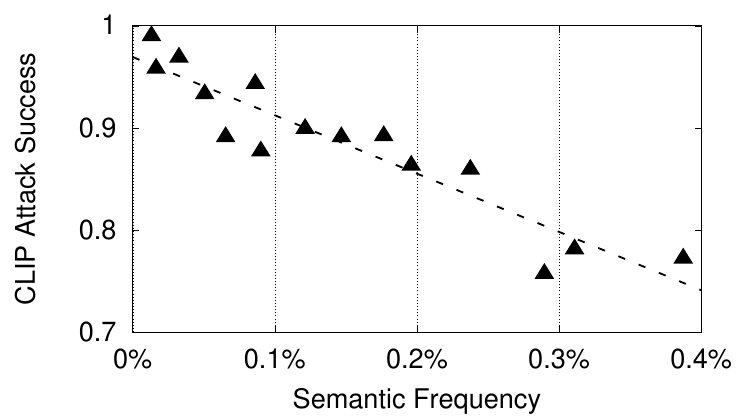}
  \vspace{-0.23in}
  \caption{Success rate of the simple poisoning attack (rated by CLIP
    classifier) correlates strongly with concept sparsity measured by semantic frequency. Results for LD-CC. Same trend
    observed on SD-V2, SD-XL, DeepFloyd.} 
  \label{fig:poison_sparsity_simple_adjusted} 
  \end{minipage}
  \hfill

\end{figure*}

Figure~\ref{fig:style-vs-object} compares the CLIP attack success rate between object and style concepts.
We observe that the simple poisoning attack is more effective at
corrupting {\em style} concepts than {\em object} concepts.  This is likely
because styles are typically conveyed visually by the entire
image, while objects define specific regions within the image.

\para{Concept Sparsity Affecting Attack Efficacy.}  
Figure~\ref{fig:poison_sparsity_simple} demonstrates 
how concept
sparsity in terms of word frequency impacts attack efficacy
and we further study the impact of semantic frequency in
Figure~\ref{fig:poison_sparsity_simple_adjusted}. For this we sample 15 object concepts with varying
sparsity levels, in terms of word and semantic frequency discussed in
\S\ref{subsec:freq}. As expected, poisoning attack is more successful when 
disrupting more sparse concepts
Moreover, semantic frequency is a more accurate representation of concept
sparsity than word frequency, because we see higher correlation between semantic frequency and attack efficacy. These empirical results confirm our hypothesis
in \S\ref{sec:hypo}.

\begin{table}[t]
  \centering
  \resizebox{0.45\textwidth}{!}{
  \centering
\begin{tabular}{rccc}
\toprule
\multirow{2}{*}{\textbf{Task}} & \multicolumn{3}{c}{\textbf{CLIP attack success rate on artist names}} \\ \cline{2-4} 
       & 100 poison & 200 poison & 300 poison \\ \midrule
LD-CC & 80\%      & 91\%       & 96\%        \\
SD-V2  & 81\%      & 94\%       & 97\%        \\
SD-XL  & 77\%      & 92\%       & 99\%        \\
DF    & 80\%      & 96\%       & 99\%        \\ \bottomrule
\end{tabular}
  }
  \vspace{-0.05in}
  \caption{Poison attack damages related concepts (artist names) when the attacker poisons given 
  art styles across $4$ generation models.  }
  \label{tab:bleed-artist-names}
\end{table}

\begin{table}[h]
  \centering
  \resizebox{0.5\textwidth}{!}{
  \centering
\begin{tabular}{rcccc}
\toprule
\multirow{2}{*}{\textbf{\begin{tabular}[c]{@{}r@{}}L2 Distance to \\ source concept(D)\end{tabular}}} &
  \multirow{2}{*}{\textbf{\begin{tabular}[c]{@{}c@{}}Average Number of \\ Concepts Included\end{tabular}}} &
  \multicolumn{3}{c}{\textbf{Average CLIP attack success rate}} \\ \cline{3-5} 
                              &      & 100 poison & 200 poison & 300 poison \\ \midrule
$D = 0$     & 1                    & 84\%       & 94\%       & 96\%        \\
$0 < D \leq 3.0$  & 5  & 81\%       & 93\%       & 96\%        \\
$3.0 < D \leq 6.0$ & 13 & 78\%       & 90\%       & 92\%       \\
$6.0 < D \leq 9.0$ & 52 & 32\%       & 41\%       & 59\%       \\
$D > 9.0$        & 1929    & 5\%       & 5\%       & 6\%       \\ \bottomrule
\end{tabular}
 }
  \vspace{-0.05in}
  \caption{Bleed through performance of the enhanced poison. (SD-XL)}
  \label{tab:bleed-through-concept-enhanced}
\end{table}

 \begin{table}[t]
  \centering
  \resizebox{0.5\textwidth}{!}{
  \centering
\begin{tabular}{rccc}
\toprule
\multirow{2}{*}{\textbf{Approach}} &
  \multirow{2}{*}{\textbf{\begin{tabular}[c]{@{}c@{}}\# of poisoned\\ concepts\end{tabular}}} &
  \multicolumn{2}{c}{\textbf{Overall model Performance}} \\ \cline{3-4} 
 &
   &
  \textbf{\begin{tabular}[c]{@{}c@{}}Alignment Score\\ (higher better)\end{tabular}} &
  \textbf{\begin{tabular}[c]{@{}c@{}}FID\\ (lower better)\end{tabular}} \\ \midrule
Clean LD-CC    & 0   & 0.31 & 17.2 \\
Poisoned LD-CC & 100 & 0.29 & 22.5 \\
Poisoned LD-CC & 250 & 0.27 & 29.3 \\
Poisoned LD-CC & 500 & 0.24 & 36.1 \\ 
Poisoned LD-CC & 1000 & 0.22 & 44.2 \\ \hline
AttnGAN        & -   & 0.26 & 35.5 \\ \hline
\begin{tabular}[c]{@{}r@{}}A model that outputs\\ random noise\end{tabular} &
  - &
  0.20 &
  49.4 \\ \bottomrule
\end{tabular}
  }
  \vspace{-0.05in}
  \caption{Overall model performance (in terms of the CLIP alignment score and FID)
  when an increasing number of concepts are being poisoned. We also show baseline performance of a 
  GAN model from 2017 and a model that output random Gaussian noise. (LD-CC)}
  \label{tab:entire-model-damage-ld}
\end{table}

\begin{figure}[t]
  \centering
  \includegraphics[width=0.9\columnwidth]{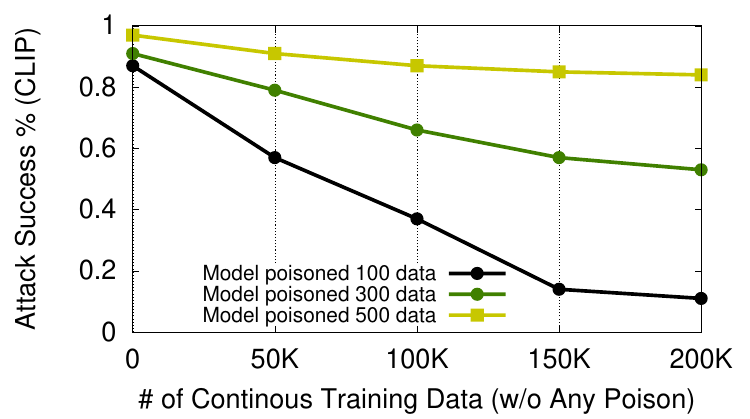}
  \caption{Nightshade's attack success rate (CLIP-based) decreases when model trainer continuously trains 
  an already-poisoned model on an increasing number of clean data. The base
  model is poisoned with 100, 300, and 500 poison data samples. }
  \label{fig:persistency}
\end{figure}

\subsection{Additional Results}
\label{subsec:bleed}

\para{Poison bleed-through. } We evaluate the ``related'' concept bleed-through effects between artists and
the art styles they are known for.
We include $195$ artists associated with $28$ styles from the Wikiart
dataset~\cite{saleh2015large}. We poison each art style $\viccon$, then test
poison's impact on generating painting of artists whose style belong to style
$\viccon$, without mentioning the poisoned style $\viccon$ in the prompt, \eg
query with ``a painting by Picasso'' for models with ``cubism'' poisoned.
Table~\ref{tab:bleed-artist-names} shows that with $200$ poison data on art
style, Nightshade achieves $> 91\%$ CLIP attack success rate on artist names
alone, similar to its performance on the poisoned art style.

\vspace{4pt}
\para{Enhancing bleed-through.} We can further enhance our poison 
attack's bleed though by broadening the sampling pool of poison text prompts:
sampling text prompts in the text semantic space of $\viccon$ rather than
with exact word match to $\viccon$. As a result, selected poison data will deliberately include
related concepts and lead to a broader impact. 
Specifically, when we calculate activation similar to the poisoned concept $\viccon$, we use
all prompts in LAION-5B dataset (does not need to include $\viccon$). Then we select top $5$K 
prompts with the highest activation, which results in poison prompts containing both $\viccon$ and
nearby concepts. We keep the rest of our poison generation
algorithm identical. This enhanced attack increases bleed through by $11\%$ in some cases
while having minimal performance degradation ($< 1\%$) on the poisoned concept 
(Table~\ref{tab:bleed-through-concept-enhanced}).

\para{Stacking multiple poisons.} Table~\ref{tab:entire-model-damage-ld} lists, for the 
LD-CC model,  the overall model
performance in terms of the CLIP alignment score and FID, when an
increased number of concepts are being poisoned.  

\final{
\para{Attack performance on complex prompts. } We also evaluate poison performance on 
longer and more complex prompts. We sample 20 prompts that contains 30+ words 
from Midjourney showcase website. We then manually replace the concept in the prompt 
to the poisoned concept (\eg  ``front view of a hyper realistic [CONCEPT] with white and blue body ...''). 
The poison attack remains successful for complex prompts ($> 80\%$ CLIP attack success rate). 
}

\newpage 

\section{Meta-Review}

The following meta-review was prepared by the program committee for the 2024
IEEE Symposium on Security and Privacy (S\&P) as part of the review process as
detailed in the call for papers.

\subsection{Summary}
The paper introduces Nightshade, a highly optimized data poisoning attack against diffusion-based text-to-image models, demonstrating the vulnerability of these models to prompt-specific poisoning attacks with a small number of optimized samples. The attack exploits concept sparsity and stealthily corrupts image generation for specific concepts, posing risks to model integrity and suggesting a tool for intellectual property protection.

\subsection{Scientific Contributions}
\begin{itemize}
\item Identifies an Impactful Vulnerability
\item Provides a Valuable Step Forward in an Established Field
\end{itemize}

\subsection{Reasons for Acceptance}
\begin{enumerate}
\item This paper illustrates a significant vulnerability in state-of-the-art models to data poisoning. The vulnerability is present even when large training datasets are used.
\item The proposed attack, Nightshade, is efficient, requiring fewer than 100 poisoned samples to significantly influence model outputs.
\item The paper is well-written, with a clear presentation and comprehensive evaluation of the proposed attack and defense mechanisms, and the experimental results are comprehensive.

\end{enumerate}

\end{document}